\documentstyle[preprint,prd,aps,epsf]{revtex}

\input{psfig}
\input{epsf.tex}

\newcommand {\Dtau}{\Delta\tau}

\begin{document}
\draft

\title{\bf
Static quark potential and string tension
for compact U(1)  in (2+1) dimensions.}

\author{Mushtaq Loan\footnote{e-mail :mushe@newt.phys.unsw.edu.au},
Michael Brunner, and Chris Hamer}

\address{School of Physics, University of New South Wales, Sydney, Australia}

\date{August 20, 2002}
\maketitle
\begin{abstract}

Compact U(1) lattice gauge theory in (2+1) dimensions is studied on 
anisotropic lattices using Standard  
Path Integral Monte Carlo techniques. We extract
the static quark potential and the string tension from 
$1.0\leq\Dtau\leq 0.333$ simulations at 
$1.0\leq \beta \leq 3.0$. Estimating the actual 
value 
of the renormalization constant, (c = 44), we observe the evidence of scaling  
in the string tension  for $1.4142\leq \beta \leq 2.5$; with the  asymptotic 
behaviour in the large-$\beta$ limit given by 
$K\sqrt{\beta}=\mbox{e}^{(-2.494\beta +2.29)}$.  
Extrapolations are made to the extreme anisotropic
or ``Hamiltonian"
limit, and comparisons are made with previous estimates 
obtained by various other methods 
in the Hamiltonian formulation.
\end{abstract}
\vspace{10mm}

\section{Introduction}

Classical Monte Carlo simulations\cite{cre79} of the path integral in Euclidean lattice
gauge theory\cite{wil74} have been very successful, and this is currently
the preferred method
for {\it ab initio} calculations in quantum chromodynamics (QCD) in the low energy
regime. Monte Carlo approaches to the Hamiltonian version of QCD
propounded by
Kogut and Susskind\cite{kog75} have been less successful, however, and lag at
least ten years
behind the Euclidean calculations. Our aim in this paper is to see whether useful
results can be obtained for the Hamiltonian version by using the standard Euclidean
Monte Carlo methods for anisotropic lattices\cite{mor97}, and extrapolating to the
Hamiltonian limit in which the time variable becomes continuous, i.e. the
lattice spacing in the
time direction goes to zero. 
The Hamiltonian version of lattice gauge theory
is less popular than the Euclidean version, but is still worthy of study. It
can provide a valuable check of the universality of the Euclidean 
results\cite{ham96},
and it allows the application of many techniques imported from quantum many-body
theory and condensed matter physics, such as strong coupling expansions\cite{ban77},
the t-expansion\cite{hor84}, the coupled-cluster method\cite{guo88}
and more recently the density matrix renormalization group\cite{byr02}
(DMRG). None of these
techniques has proved as useful as Monte Carlo in (3+1) dimensions; but in
lower
dimensions they are more competitive.

A number of Quantum Monte Carlo methods have been applied to Hamiltonian
lattice gauge theory in the past, with somewhat mixed results. A 
``Projector Monte Carlo"
approach\cite{bla83,deg85} using a strong coupling representation for the gauge
fields and a Greens Function Monte Carlo (GFMC) approach
\cite{chi84,hey83}, which makes use of a weak coupling representation of 
the gauge fields. However both the approaches run into difficulties in 
that the former approach runs into 
difficulties for non-Abelian models, 
and the later requires the use of a ``trial wave function" to
guide random walkers in the ensemble towards the preferred regions of
configuration space\cite{kal66}. This introduces a variational element into the
procedure, in that the results may exhibit a systematic dependence on the trial
wave function \cite{sam99,ham00,ham00a,liu74,whi79}.
For this reason, we are forced to look
yet again for an alternative approach.

As mentioned above our aim in
this paper is to use
standard Euclidean path integral Monte Carlo 
(PIMC) techniques for anisotropic lattices, and
see whether useful results can be obtained in the Hamiltonian limit.
Morningstar and Peardon\cite{mor97} showed some time ago that the use of
anisotropic lattices can be advantageous in any case, particularly for
the measurement of glueball masses. We use a number of their techniques
in what follows. 

As a first trial of this approach, we treat the U(1) gauge model in
(2+1)D, which is one of the simplest models with dynamical gauge degrees
of freedom, and has also been studied extensively by other means (see
Section II). Path integral Monte Carlo (PIMC) methods were applied to this
model a long time ago by Hey and collaborators\cite{amb82,cod86}, but
the techniques used at that time were not very sophisticated, and the
results were rather qualitative. Very little has been done since then
using this approach on this particular model.

The rest of this paper is organised as follows. In section II we discuss
the U(1) model in (2+1) dimensions in its lattice formulation, and
outline some of the work done on it previously. The details
of the simulations, including the generation of the gauge-field
configurations, the
construction of the Wilson loop operators,  and the
extraction of the potential and string tension
estimates are described in section III.
In section IV we present
our main results for the mean plaquette, static quark potential, string
tension.  The static
quark potential has not previously been exhibited for this model, as far
as we are aware. Finally
we make an extrapolation to the Hamiltonian limit, and comparisons are
made with estimates obtained by other means in that limit. (Sec. 1V).
We find that indeed the PIMC method can give better results than other
methods, even in the Hamiltonian limit.
Our conclusions are summarized in Sec. VI.

\section{THE U(1) MODEL}

Consider the  isotropic Abelian U(1) lattice gauge theory in three dimensions.
 The theory is
defined by the action\cite{wil74}
\begin{equation}
S = \beta
\sum_{r,\mu,\nu}\mbox{ReTr}P_{\mu\nu}
\label{eqn1}
\end{equation}
where
\begin{equation}
\mbox{P}_{\mu\nu}(r) =\left[1-
Re Tr\{\mbox{U}_{\mu}(r)\mbox{U}_{\nu}(r+\hat{\mu})
\mbox{U}_{\mu}^{\dagger}(r+\hat{\nu})\mbox{U}_{\nu}^{\dagger}(r)\right\}]
\label{eqn2}
\end{equation}
is the plaquette variable given by the product of the link variables taken around an
elementary plaquette. The link variable  $\mbox{U}_{\mu}(r)$ is defined by
 \begin{equation}
 \mbox{U}_{\mu}(r) =\mbox{exp}[ieaA_{\mu}(r)] =
\mbox{exp}[i\theta_{\mu}(r)]
\label{eqn3}
 \end{equation}
 where in the compact form of the model, $\theta_{\mu}(r)
 (=eaA_{\mu}(r)) \in [0,2\pi]$
 represents the gauge field  on the directed link $r \rightarrow r+\hat{\mu}$.
 The parameter $\beta$ is related to the bare gauge coupling by
 \begin{equation}
 \beta =\frac{1}{g^{2}}
\label{eqn4}
 \end{equation}
 where $g^{2} = ae^{2}$, in (2+1) dimensions.

The lattice U(1) model in (2+1) dimensions has been studied by many authors, and possesses some important
similarities with QCD (for a more extensive review, see for example ref.
\cite{ham94}). If one takes the ``naive" continuum limit  at a fixed
energy scale, one regains the simple continuum theory of non-interacting
photons\cite{gro83}; but if one renormalizes or rescales in the standard
way so as to maintain the
mass gap constant, then one obtains a confining theory of free massive bosons.
Polyakov\cite{pol78} showed that a linear potential appears between
static charges due to instantons in the lattice theory; and
G{\" o}pfert and Mack\cite{gop82} proved that in the continuum limit the
theory converses to a scalar free field theory of massive bosons.
They found that in that limit the mass gap behaves as
\begin{equation}
am_{D} = \sqrt{\frac{8\pi^{2}}{g^{2}}}\mbox{exp}\bigg(-\frac{\pi^{2}}{g^{2}}
v(0)\bigg)
\label{eqn05}
\end{equation}
while the string tension is bounded by
\begin{equation}
a^{2}\sigma = c\sqrt{\frac{g^{2}}{2\pi^{2}}}\mbox{exp}(-\frac{\pi^{2}}{g^{2}}v(0))
\label{eqn06}
\end{equation}
where $v(0)$ is the Coulomb potential at zero distance, and has a value
in lattice units
\begin{equation}
v(0) = 0.2527
\label{eqn07}
\end{equation}
for the isotropic case. They argue that eq. (\ref{eqn06}) represents the 
true
asymptotic behaviour of the string tension, where the constant $c$ is
equal to $8$ in classical approximation. The theory has a non-vanishing 
string tension for arbitrary large $\beta$, similiar to behaviour  
expected for the string tension in non-abelian lattice gauge theories in 
4-dimensions.

In the small-$\beta$ limit the string 
tension behaves as:
\begin{equation}
a^{2}\sigma = -\mbox{ln}\beta /2 + ....
\label{eqn08}
\end{equation}
and the large-$\beta$ behaviour is given by eq. (\ref{eqn06}).

Similar   
behaviour will apply in the anisotropic case. Generalizing
discussions by Banks {\it et al}\cite{ban77} and
Ben-Menahem\cite{ben79}, we find (see the appendix A) that the exponential
factor takes exactly the same form in the anisotropic case, with a
Coulomb potential $v(0)$ changing to $0.3214$ in the Hamiltonian limit.

 For an anisotropic lattice, the gauge action becomes\cite{kar82,bur88}
 \begin{equation}
 S = \beta_{s}\sum_{r,i<j}P_{ij}(r) +\sum_{r,i} \beta_{t}P_{it}(r)
\label{eqn09}
 \end{equation}
 where
  $P_{ij}$ and $P_{it}$ are the spatial and temporal plaquette
variables respectively.
 In the classical limit
 \begin{eqnarray}
 \beta_{s} & = & \frac{a_{t}}{e^{2}a_{s}^{2}}  = \frac{1}{g^{2}}\Dtau 
\nonumber\\
 \beta_{t} & = & \frac{1}{e^{2}a_{t}}  = \frac{1}{g^{2}}\frac{1}{\Dtau}
\label{eqn10} 
\end{eqnarray}
 where $\Dtau = a_{t}/a_{s} $ is the anisotropy
parameter, $a_{s}$ is the lattice
 spacing in the space direction, and $a_{t}$
is the temporal spacing.
 The above action can be written as
 \begin{eqnarray}
 S & =& \beta \bigg[\Dtau \sum_{r}\sum_{i<j}\bigg(1-\cos \theta_{ij}(r)\bigg)
+\frac{1}{\Dtau }\sum_{r,i}\bigg(1-\cos \theta_{it}(r)\bigg)\bigg]
\label{eqn11}
\end{eqnarray}

In the limit $\Delta \tau
 \rightarrow 0$, the time variable becomes
continuous, and
 we obtain the Hamiltonian limit of the model (modulo a Wick rotation back to Minkowski space).
The Hamiltonian version of the model has been studied by many methods:
some recent studies include series expansions
\cite{ham92,ham96}, finite-lattice techniques\cite{irv83}, the t-expansion
\cite{hor87,mor92}, and coupled-cluster techniques\cite{dab91,fan96}, as well
as Quantum Monte Carlo methods\cite{chi86,koo86,yun86,ham94}.
Quite accurate estimates have been obtained for the string
tension and the mass gaps, which can be used as comparison for our
present results. The finite-size scaling
properties of the model can be predicted using an effective Lagrangian
approach combined with a weak-coupling expansion
\cite{ham93}, and the predictions agree very well with finite-lattice data
\cite{ham94}.

Some analytic results for the model in the weak-coupling approximation appear
in the appendix.

\section{METHODS}

\subsection{Path Integral Monte Carlo algorithm}

We perform
standard path integral
Monte Carlo simulations on a finite lattice of size $N_{s}^{2}\times
N_{\tau}$, where
$N_{s}$ is the number of lattice sites in the space direction and
$N_{\tau}$
in the temporal direction, with spacing ratio
$\Delta \tau = a_{t}/a_{s} $.
By varying $\Delta \tau $ it is possible to change $a_{t}$, while
keeping the spacing in the spatial direction fixed.
The
simulations were performed on lattices with $N_s=16$ sites in each of
the two
spatial directions and
$N_{t}= 16 - 64$ in the temporal direction for a range
of couplings $\beta = 1 - 3$.

The ensembles of field configurations were generated by using a Metropolis
algorithm. Starting from an arbitrary initial configuration of link angles, we
successively update
link angles $\theta _{\mu}(\vec{r},\tau)$
at positions $(\vec{r},\tau)$ which are chosen randomly each time. We propose a  change $\Delta\theta$
to  this link angle, which is randomly drawn from a uniform
distribution on $[-\Delta,\Delta]$, where $\Delta$ is adjusted for
each set of parameters to give an acceptable ``hit rate" around 70-80\%.
The change is accepted or rejected
according to the standard Metropolis procedure.

For high anisotropy ($\Delta \tau <<1$), any change in a
time-like plaquette will produce a large change in the action, whereas changes to the
space-like plaquettes will cause a much smaller change in the action.
This makes the system very ``stiff" against variations in the time-like
plaquettes, and therefore very slow to equilibrate, with long
autocorrelation times.
To alleviate this problem, we used a Fourier update
procedure\cite{bat85,dav90}. Here proposed changes are made to space-like links which
are designed to alter space-like plaquette values much more than the
time-like plaquette values. At randomly chosen intervals and random
locations,
we propose a non-local change
$\Delta\theta (\vec{r},\tau)= X\sin k(\tau -\tau_{0} )$ on a ``ladder"
of
space-like links extending half a wavelength $(\lambda = 2\pi /k)$ in the
time direction, where both $k$ and $X$ are randomly chosen at each update 
from uniform
distributions in $[0,\pi/2]$.
We replaced approximately $30\%$ of the ordinary
Metropolis updates with Fourier updates for anisotropy $\Delta \tau
>0.444 $ and $50\%$ for highly anisotropic cases ($\Delta \tau
<0.444 $). These moves satisfy the requirements of detailed balance and
ergodicity for the algorithm.

\begin{figure}
\centerline{\psfig{file={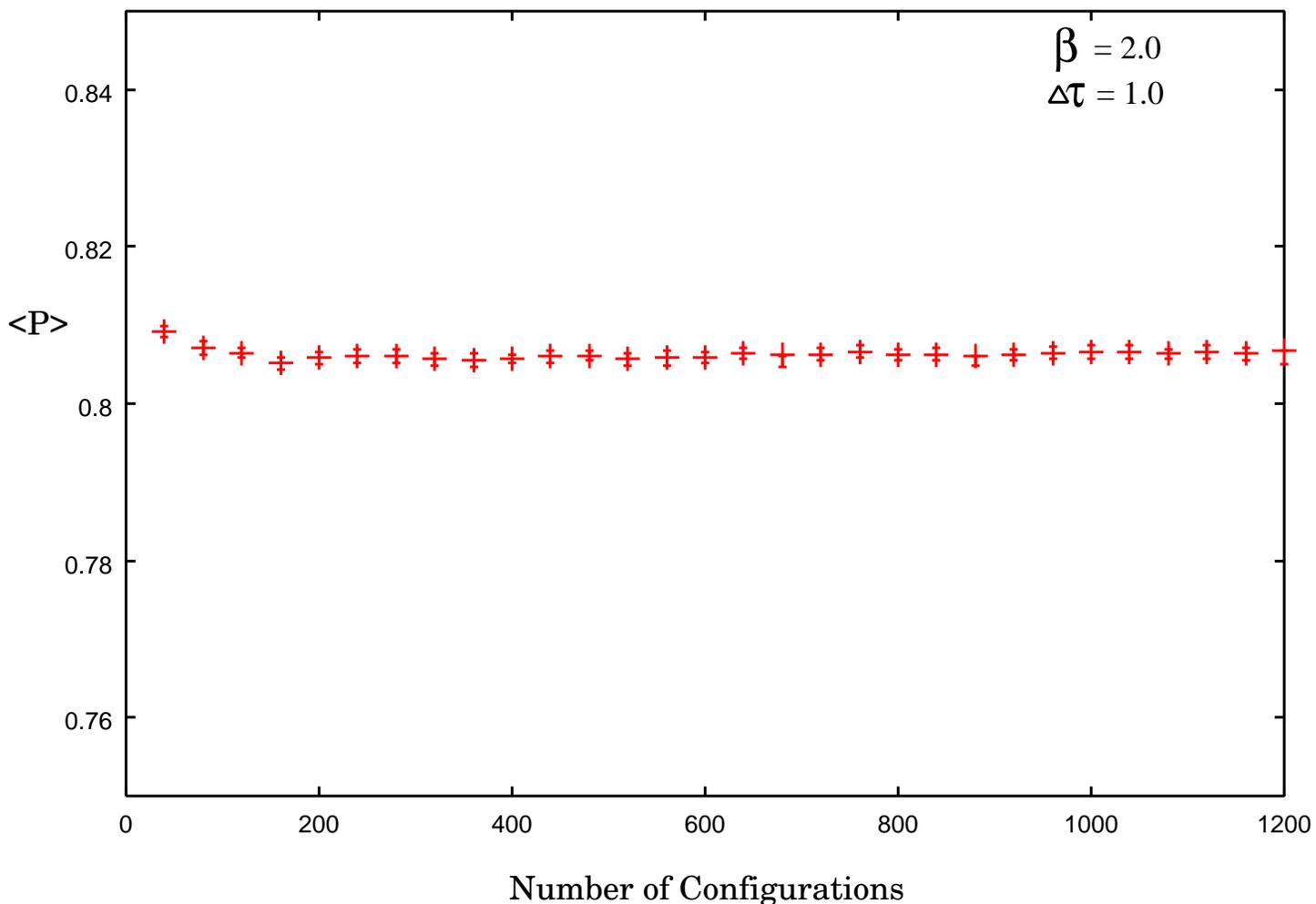}}}
\caption{Plot of plaquette average against the number of configurations. 
The loop average is seen to reach its  equilibrum value, after discarding the 
initial number of sweeps (50000 for $16^{3}$ lattice).}
\label{fig1}
\end{figure}

A single
sweep involves attempting $N$ changes to randomly chosen links of the
lattice, where $N (=3N_{t}N_s^{2})$ is the total number of links on the
lattice. The first several thousand sweeps are discarded to allow the system to
relax to equilibrium.
Fig. \ref{fig1}
shows measurements of the mean plaquette
for $\beta = 2.0$ and $\Delta \tau =1.0 $, and it can be seen that
equilibrium is reached after
about 50,000 thousand sweeps, with the
measurements fluctuating about the equilibrum value thereafter.
For highly anisotropic cases the system was much slower to equilibrate,
despite the Fourier acceleration, and in the worst case the
equilibration time was of order 100,000 thousand sweeps.

After discarding the initial sweeps, the
configurations were stored every 250 sweeps
thereafter  for later analysis.
Ensembles of about 1,000 configurations were stored  to
measure the static quark potential and glueball masses
 at each $\beta$ for $\Delta \tau \geq 0.4$,
and 1,400 configurations for $\Delta \tau \leq 0.333$.
 Measurements made on
these stored configurations were grouped into 5 blocks, and then the
mean and standard deviation of the final quantities were estimated by
averaging over the ``block averages", treated as independent
measurements. Each block average thus comprised 50,000 - 70,000 sweeps.

\subsection{Interquark Potential}

The static quark-antiquark potential, $V(\bf {r})$ for various spatial
separations $\bf{r}$ is extracted from the expectation values of the
Wilson loops.
 The timelike Wilson
loops  are expected to behave as:
 \begin{equation}
W({\bf r},\tau) \simeq Z({\bf r})\mbox{exp}\big[-\tau V({\bf r})] + $(excited state contributions)$
\label{eqn12}
\end{equation}
We have averaged only over loops $({\bf x_{0}},\tau_{0}) \rightarrow
({\bf x_{0}+r},\tau_{0}) \rightarrow ({\bf x_{0}+r},\tau_{0}+\tau)
\rightarrow ({\bf x_{0}},\tau_{0}+\tau) \rightarrow ({\bf
x_{0}},\tau_{0})$ which follow either two
sides of a rectangle between ${\bf x_{0}}$ and ${\bf x_{0}+r}$ or a
single-step `staircase' route, to estimate $W({\bf r},\tau)$.
To suppress the excited state
contributions,
a simple APE smearing technique\cite{alb87,tep86} was used on the space-like
variables. In this technique an iterative smearing procedure is used to
construct Wilson loop (and glueball) operators with a very high degree of
overlap with the lowest-lying state.
In our single-link smoothing procedure, we replace every space-like link
variable by
\begin{equation}
U_{i} \rightarrow P\left[\alpha U_{i} +
\frac{(1-\alpha)}{2}\sum_{s}U_{s}\right]
\label{eqn13}
\end{equation}
where the sum over ``s" refers to the ``staples", or 3-link paths bracketing the given link on either side in the spatial plane, and P denotes a
projection onto the group U(1), achieved by renormalizing the magnitude to unity.  We used a smearing parameter $\alpha = 0.7$
and up to 10 iterations of the smearing process.

To further reduce the statistical errors, the timelike
Wilson loops were constructed from
`` thermally averaged" temporal links\cite{par83}. That is, the temporal links $U_t$
in each Wilson loop were replaced by their thermal averages
\begin{equation}
\bar{U_t} = \int dU U \exp(-S[U])/ \int dU \exp(-S[U])
\label{eqn14}
\end{equation}
where the integration is done over the one link only, and depends on the
neighbouring links. For the U(1) model, the result can easily be
computed in terms of Bessel functions involving the ``staples" adjacent
to the link in question. This was done for all temporal links except
those adjacent to the spatial legs of the loop, which are not
``independent"\cite{par83}. The procedure has a dramatic effect in
reducing the statistical noise, by up to an order of
magnitude 10, which corresponds to a factor 100 in the 
statistics\cite{mor97}.

\begin{figure}
\centerline{\psfig{file={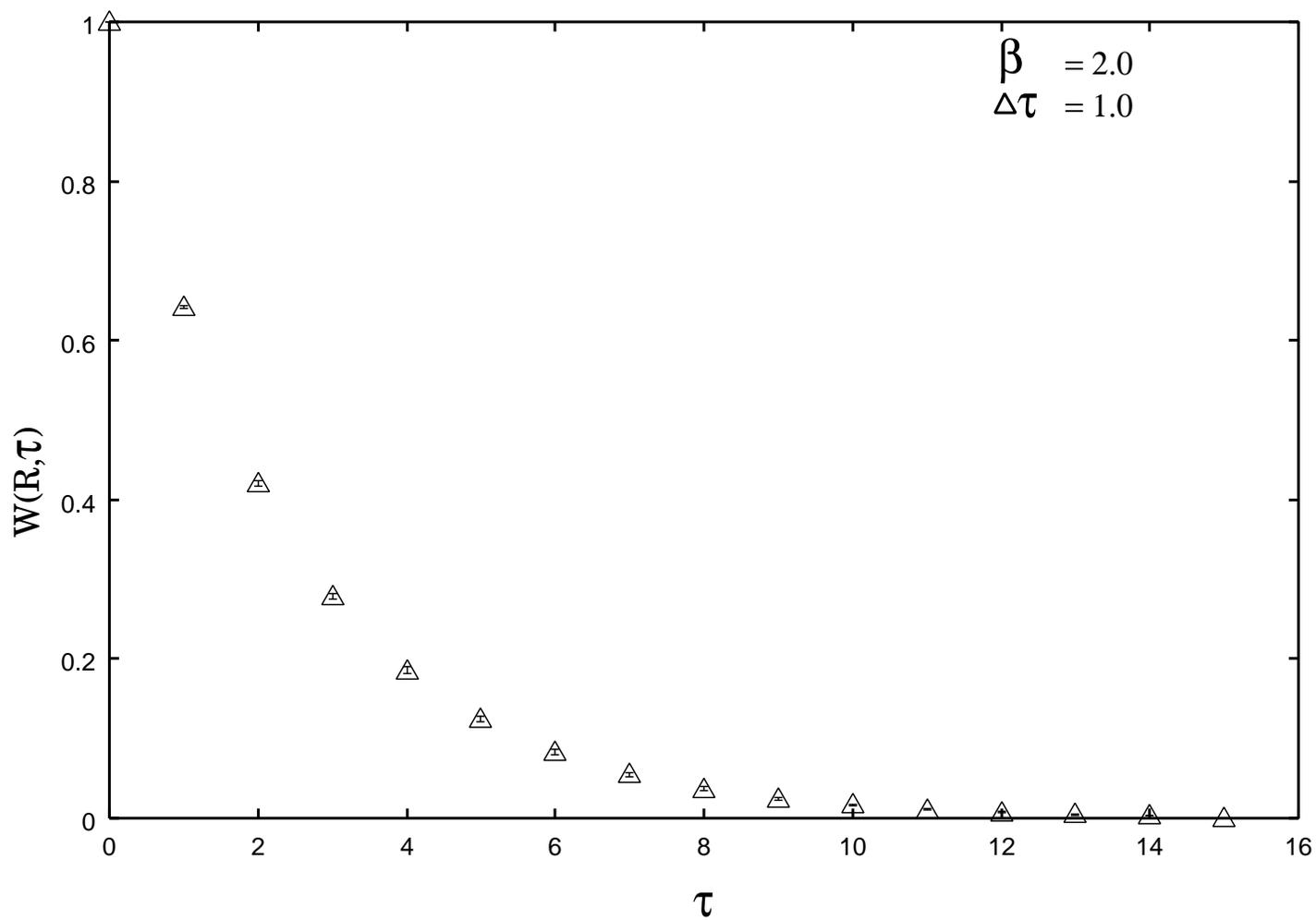}}}
\caption{Plot of a $4\times 1$ Wilson loop against $\tau$ from $\Delta 
\tau =1.0$ simulations for $\beta =2.0$}
\label{fig2}
\end{figure}

\begin{figure}
\centerline{\psfig{file={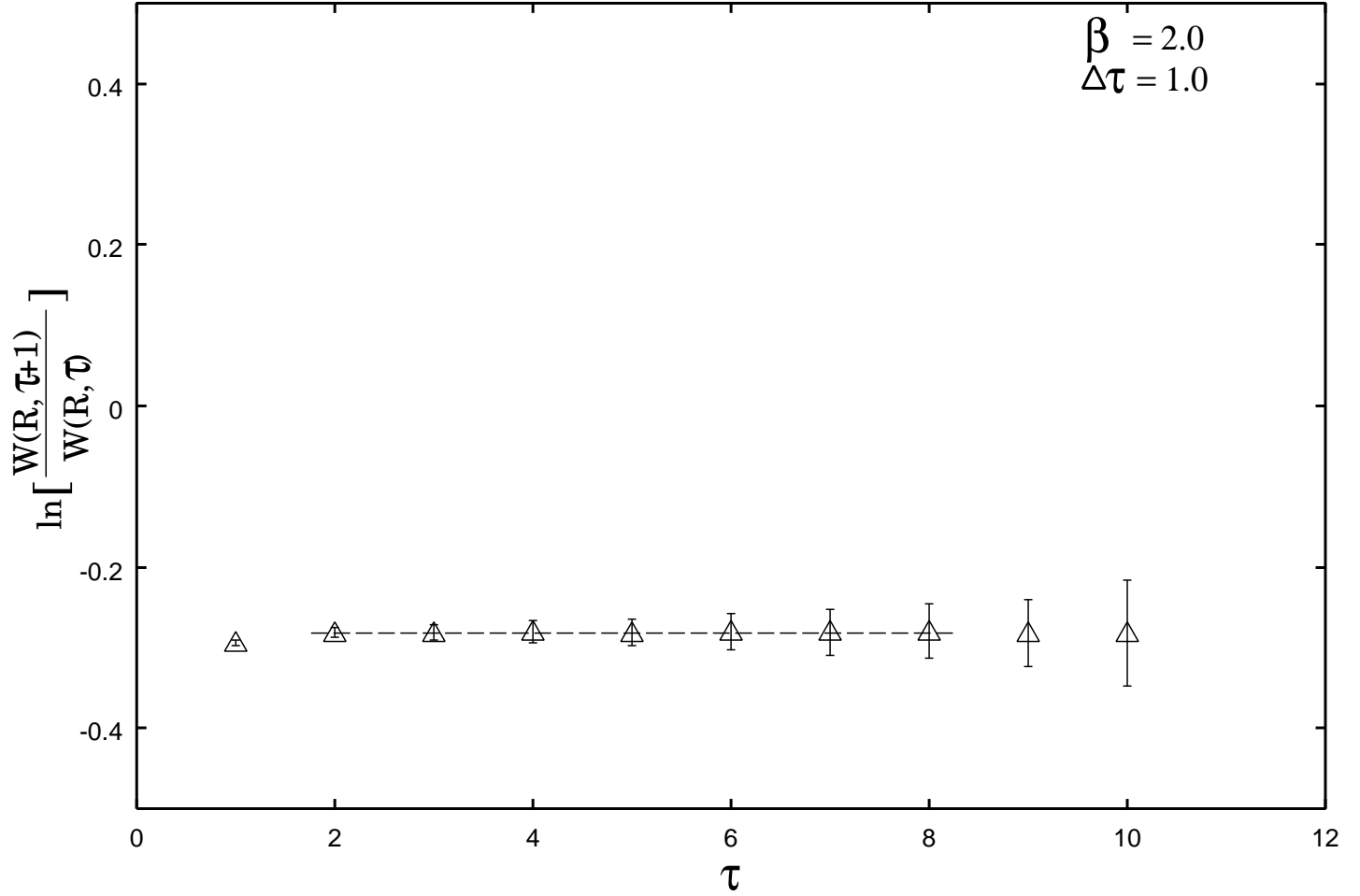}}}
\caption{Ratio of the Wilson loops as a function of $\tau$ for fixed $x$ 
and $y$ from $\beta = 2.0$ and $\Delta \tau =1.0$ simulations. The ratio 
is seen reaching  its plateau at very  early times  and is very close to  
its asymptotic large-$t$ plateau.}
\label{fig3}
\end{figure}

Fig. \ref{fig2}  shows an example of the exponential decrease of the 
$4\times 1$Wilson  loop
$W(\vec{r},\tau)$ with $\tau$.
Estimates of the potential can now be found from the ratios of successive
 loop values:
\begin{equation}
V({\bf r}) = \frac{1}{\Delta \tau}\mbox{ln}
\bigg[\frac{W({\bf r},\tau+\Delta \tau)}{W({\bf r},\tau)}\bigg]
\label{eqn15}
\end{equation}
This ratio is expected to be independent of $\tau$ for $\tau >>0$.
Fig.\ref{fig3} shows the effective potential as a function of Euclidean
time, $\tau$, obtained from $\Delta \tau =1.0$ at $\beta =2.0$
for $r =4$ simulation. We find that ratio reaches its plateau
behavior at very early Euclidean times, which reflects a good optimized
smearing. We used $\tau =1$ to make our estimates of $V(\bf{r})$. 

\section{Simulation results at finite temperature}

Simulations were carried out for lattices of $ N_s^{2}\times N_{t}$ sites,
with $N_s=16$ and $N_{t}$ ranging from 16 to 48
 sites, with periodic boundary conditions. Each
run involved 250,000
sweeps (350,000 for high anisotropy) of the lattice, with 60,000 sweeps
(100,000 high anisotropy) discarded to
 allow for equilibrum, and configurations recorded every 250  sweeps
thereafter. Coupling values from $\beta = 1.0$ to
 3.0 were explored at anisotropies $\Delta \tau$
ranging from 1 to 1/3. We fixed $\Dtau =16/N_t$ in the first pass, so that 
the lattice size
$T=N_ta_t$ in the time direction remains constant, corresponding to a fixed (low)
temperature.

\subsection{Mean Plaquette}
\begin{figure}
\centerline{\psfig{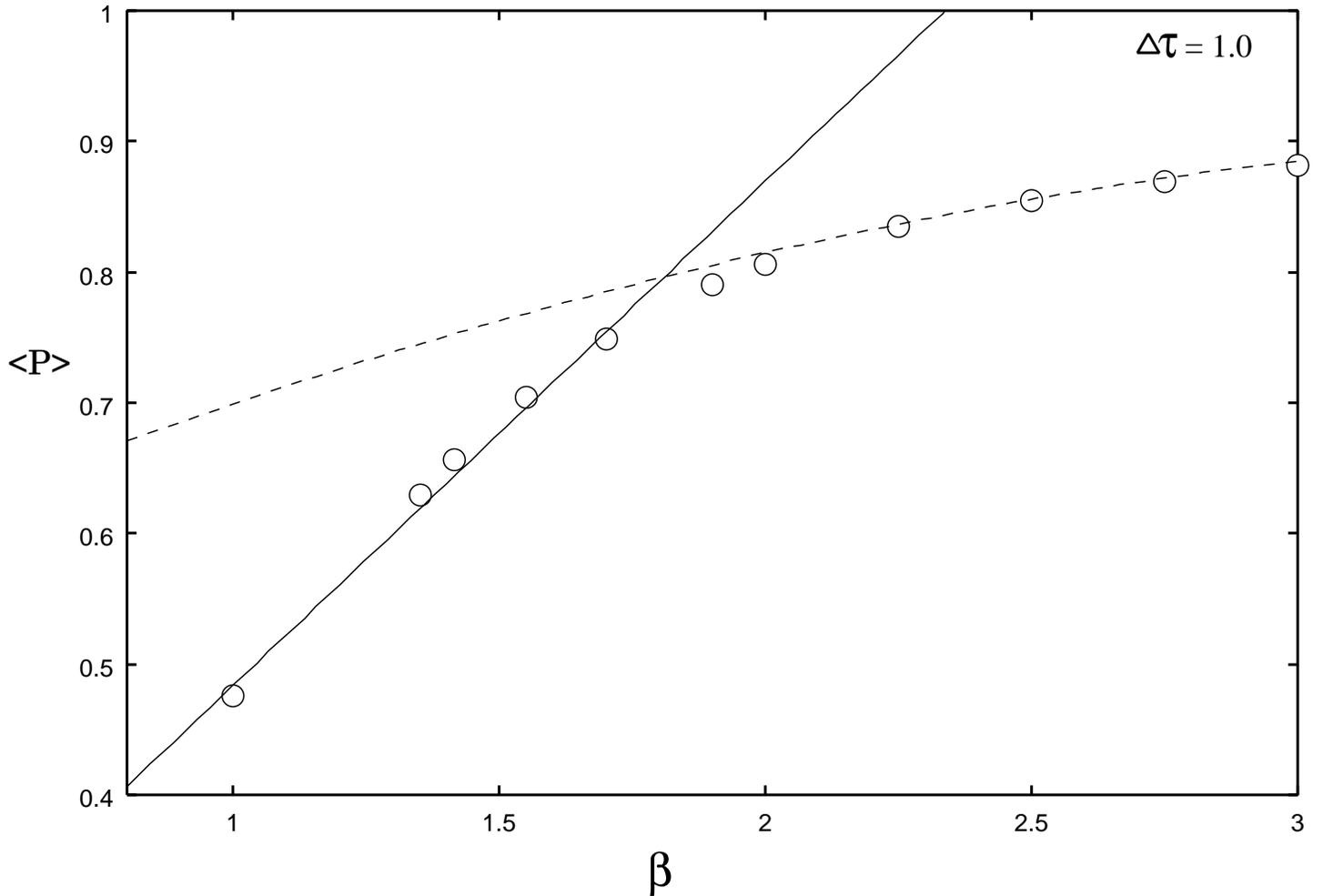}}
\caption{Plot of average plaquette 
 as a function of $\beta$. The $\circ$ symbols label  our Monte
Carlo estimates. The solid curve represents the $O(\beta^{4})$ 
strong-coupling 
expansion [49]  and the dashed curve represents the $O(1/\beta^{5})$ 
weak-coupling expansion [50].}
\label{fig4}
\end{figure}

\begin{figure}
\centerline{\psfig{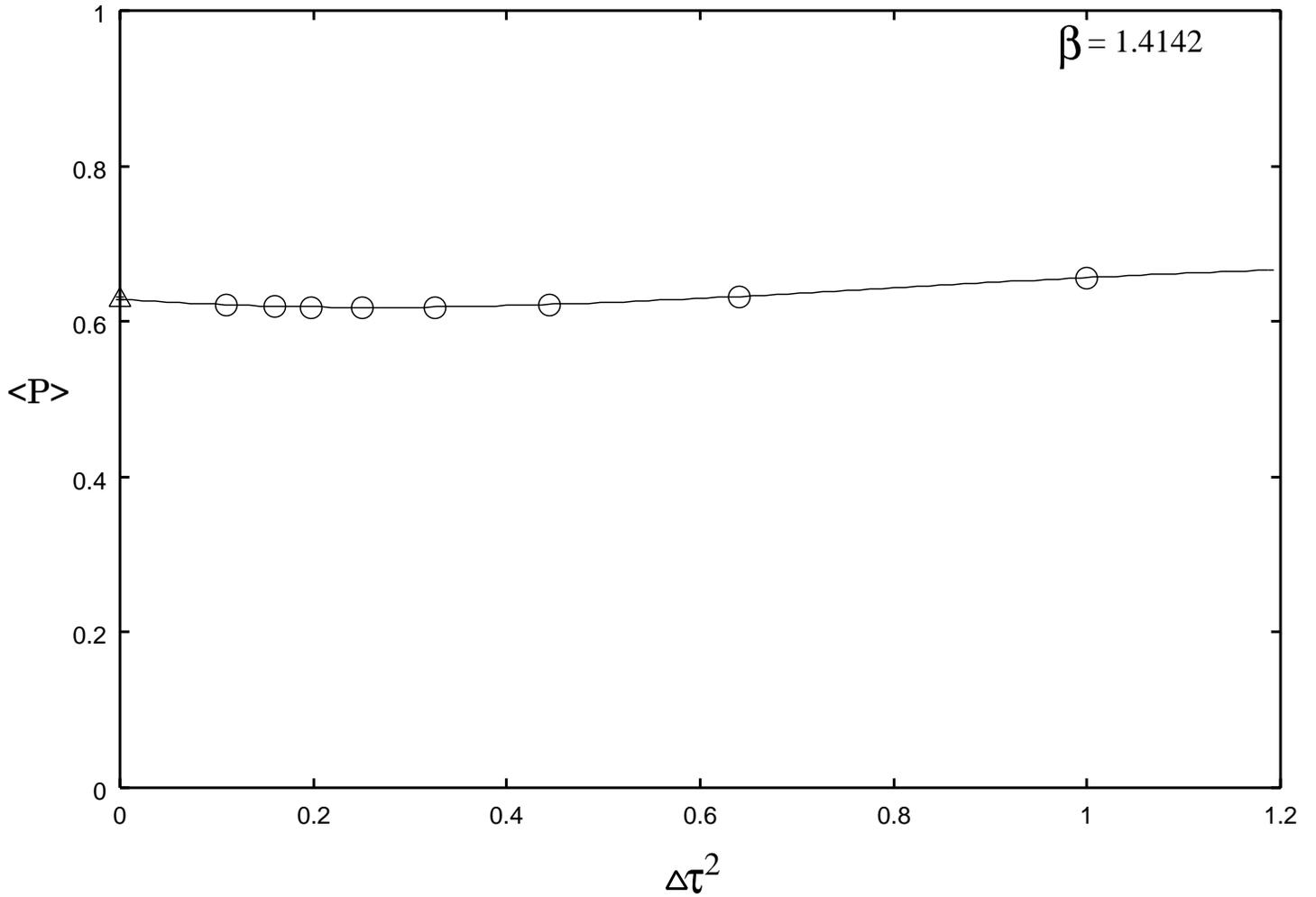}}
\caption{The mean plaquette  as a function of $\Delta \tau^{2}$ at
$\beta = 1.4142$. The solid curve is cubic fit, in powers of $\Delta 
\tau^{2}$, to the data extrapolated to the Hamiltonian limit. 
$\triangle$ symbol represents the series estimates [27] in that limit.}
\label{fig5}
\end{figure}

\begin{figure}
\centerline{\psfig{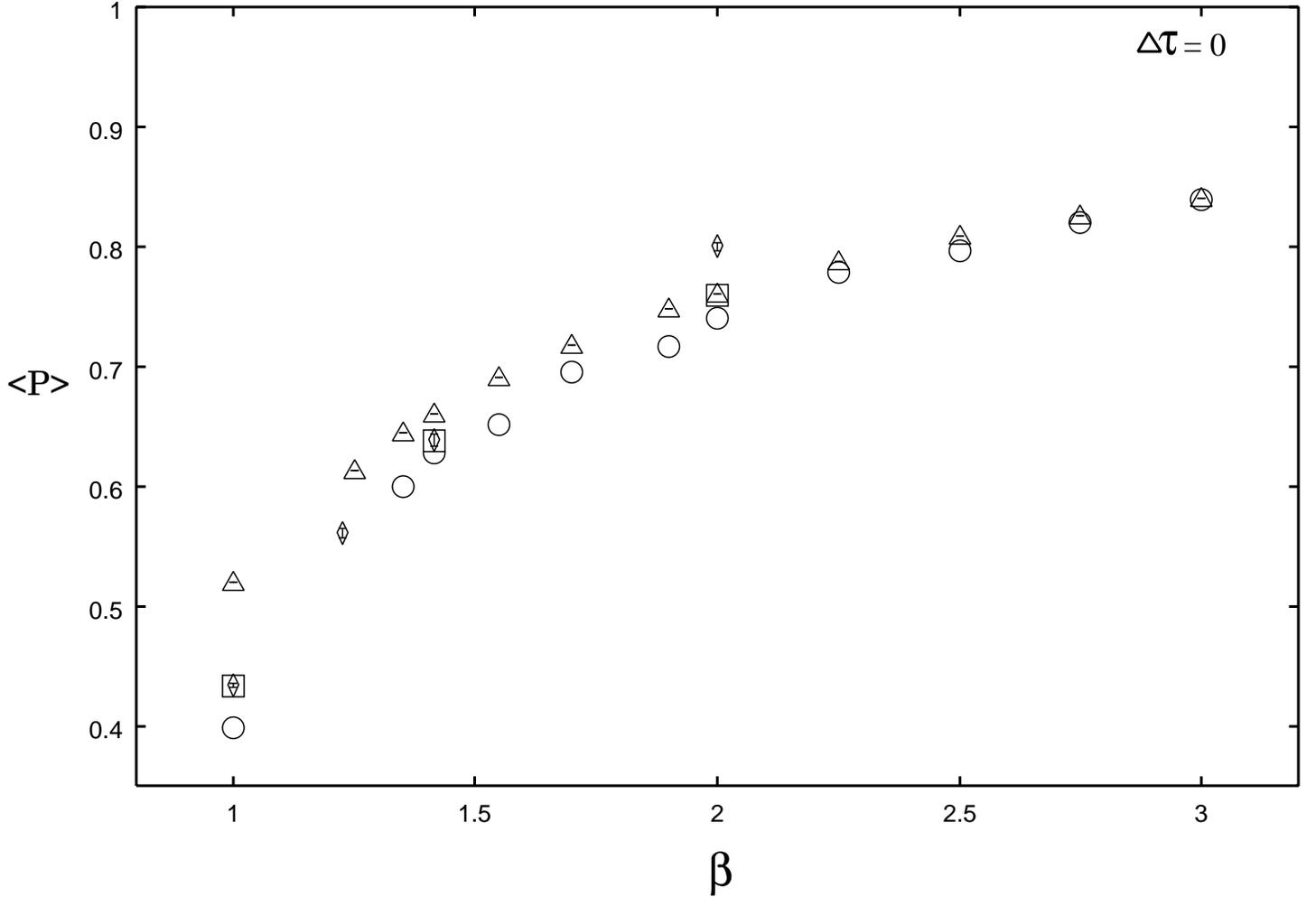}}
\caption{Mean plaquette estimates  against $\beta$ at $\Delta \tau =0$.
The earlier estimates from strong-coupling expansion and  Greens function
Monte Carlo are represented by 
$\Diamond$  and $\Box$ symbols 
respectively.  Our Monte Carlo estimates and 
 weak-coupling expansion results, (Eq.B31),
are represented by $\bigcirc$ and $\triangle$ respectively.} 
\label{fig6}
\end{figure}

The mean plaquette, or equivalently the $1\times 1$ Wilson loop, was 
measured for various $16\times 16\times N_{t}$ (with $N_{t}$ = 16 up to 48) 
sizes and for a variety of $\beta$ values, ranging from strong coupling to 
weak coupling regime. 
Fig. \ref{fig4} shows the behaviour of the mean spatial plaquette $<P>$ 
for different
$\beta $, at fixed $\Dtau=1$ (isotropic case).
At small coupling (large $\beta $) the plaquette expectation
value is
expected to approach 1, and conversely it approaches 0 in the high
coupling (small $\beta $) limit. The plot also shows the 
$O(\beta^{4})$
strong-coupling \cite{bhanot80} and $O(1/\beta^{5})$ weak coupling
\cite{hors81} 
expansion.
The dashed curve represents the weak-coupling series
and strong-coupling series is represented by the solid 
curve. As can be seen that the data follow closely the weak coupling 
expansion down to $\beta=2.0$ and strong coupling expansion upto 
$\beta = 1.70$ and the 
results agree very well. The variation of $<P>$ with coupling is extremely 
smooth, with no sign of phase transition, as we should expect. The 
cross-over from strong to weak coupling seems, from the Monte Carlo data, 
to take place quite raidly in the region
$\beta \approx 1.8 - 2.0$.
 Horsley and Wolff \cite{hors81} 
investigates the effect of finite size lattice in their 
weak-coupling expansion calculations. They find that such effects enter at 
order $1/\beta^{2}$ as a correction of order $1/l^{D}$, where $l$ is the 
lattice size and $D$ is the number of dimensions.  Thus for U(1) in (2+1) 
and  $l>10$, finite size effects should be essentially 
negligible.\\
Fig. \ref{fig5} shows a plot of our estimates
of $<P>$ as a function of anisotropy ${\Delta \tau}^{2}$ for the case
$\beta = \sqrt{2}$.
We would like to make contact with previous Hamiltonian studies
by showing that the mean plaquette value approaches
to previously known values in the Hamiltonian limit
 $\Delta \tau \rightarrow 0$.
The extrapolation
was performed using a simple cubic fit in powers of $\Dtau^{2}$.
In this limit our results agree very well with the
Hamiltonian estimate obtained by Hamer et al\cite{ham00}.
The estimates of the mean plaquette $<P>$ in the Hamiltonian limit are 
graphed as a function of $\beta$ in Fig.\ref{fig6} and compared with the 
series estimates at strong and weak coupling \cite{ham92}, 
 and  Greens 
function Monte Carlo results\cite{ham00}. 
At large $\beta$, the data is in good agreement with 
predictions of weak-coupling 
perturbation theory. At  small $\beta$, the data follow closely 
 the strong-coupling and Greens function Monte Carlo estimates, 
except  at very small $\beta$.
We attribute the  discrepnacy  to the finite-temperature 
effects  and a better agreement is expected  at 
zero-temperature [52].
\subsection{Static quark potential and string tension}
\begin{figure}
\centerline{\psfig{file={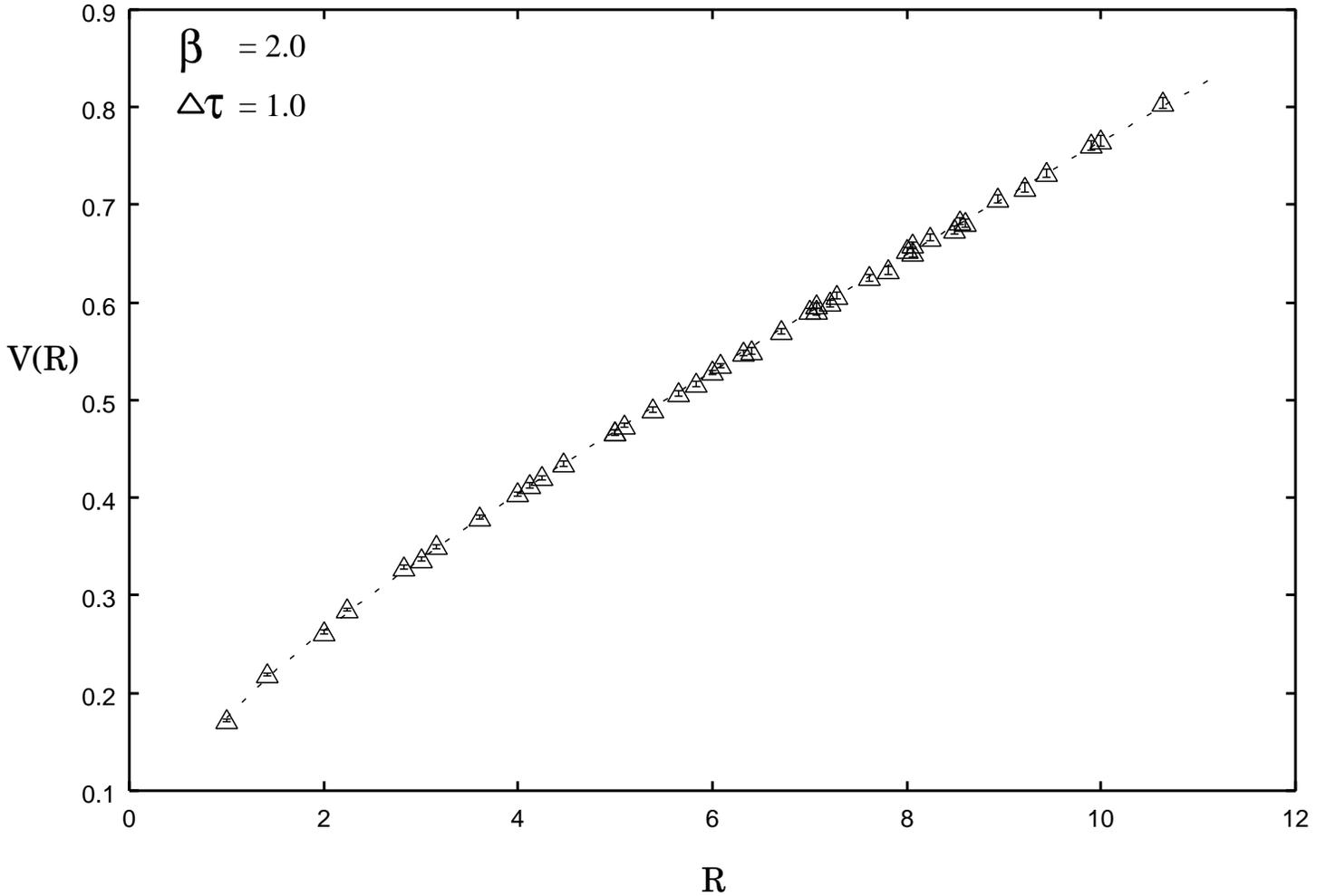}}}
\caption{The static-quark potential, $V(R)$ as a function of  
the separation R. 
This plot includes the measurements from $\beta=2.0$ for $\Delta \tau
= 1.0$ with 10 sweeps of smearing at smearing parameter $\alpha = 0.7$.
The analysis errors are smaller than the plot symbols. The dashed 
line is the fit to the form $V(R)=a+bR+c\mbox{log}(R)$.}
\label{fig7}
\end{figure}

The effective potential is obtained from the ratio of successive Wilson 
loop values:
\begin{equation}
V_{\tau}({\bf{r}})=\mbox{ln}\bigg(\frac{W({\bf{r}},\tau +\Delta \tau)}{
W({\bf{r}},\tau)}\bigg)
\label{eqn26}
\end{equation}
Fig. \ref{fig7} shows a  graph of the static quark potential V(R)   as a
function of radius R at $\beta = 2.0$ and $\Delta \tau = 1.0$.
To extract the string tension, the curve is well fitted by
a form
  \begin{equation}
  V(R) = a +b\ln R+\sigma R,
\label{eqn27}
  \end{equation}
including a logarithmic Coulomb term as expected for classical
QED in (2+1) dimensions which dominates the behaviour at small distances,
and
a linear term as predicted by Polyakov\cite{pol78} and G{\" o}pfert and
Mack\cite{gop82} dominating the behaviour at large distances. The linear
behaviour at large distances is very clear, but the data do not extend
to very small distances, so there is no real test of the presumed
logarithmic behaviour in this regime.

\begin{figure}
\centerline{\psfig{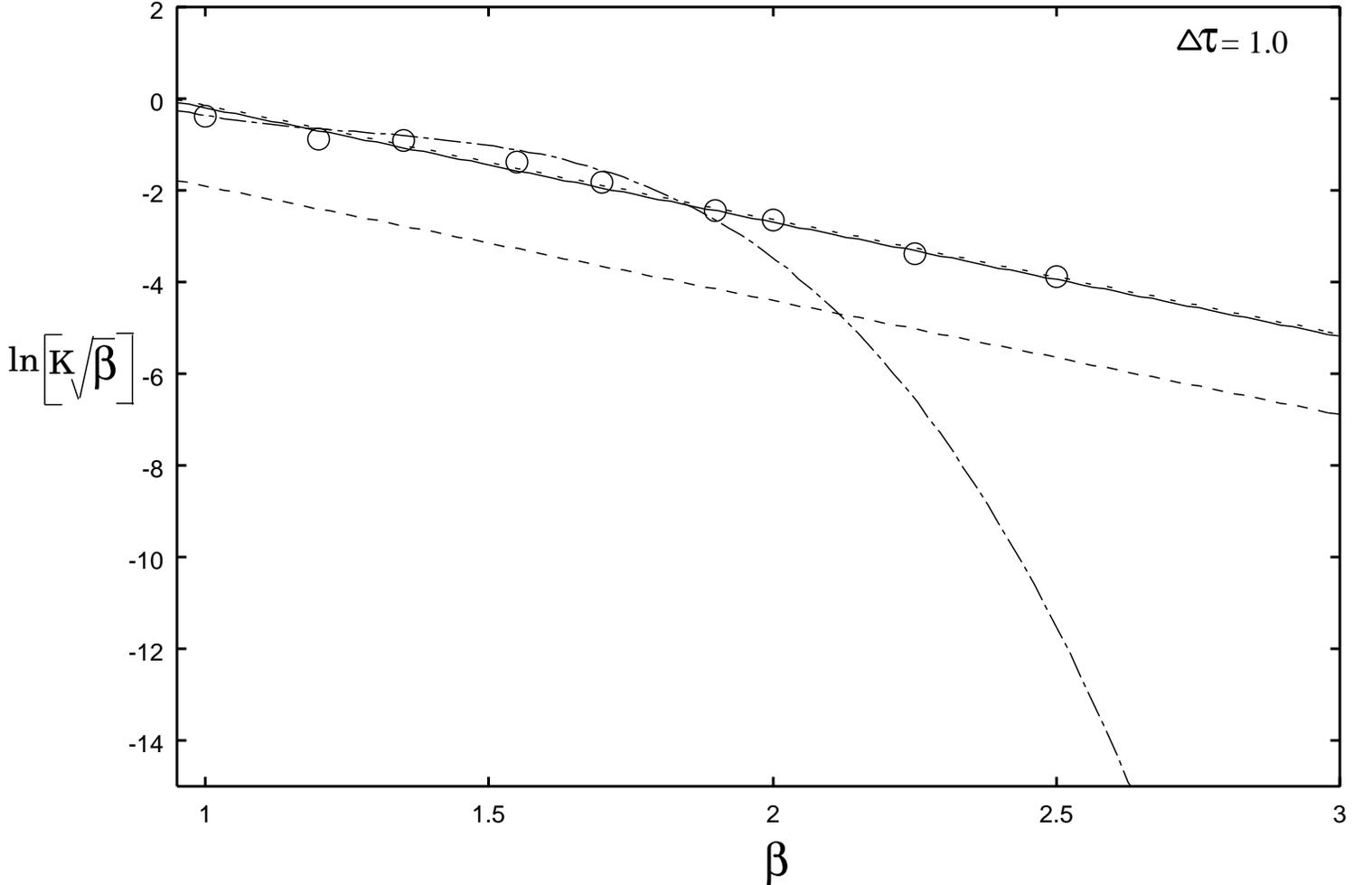}}
\caption{String tension as a function of $\beta$. The solid curve 
represents the  
weak-coupling result  using Villian 
approximation,  
with our estimated value for renormalization constant, $c$(=44) 
and the dotted curve coinciding with the solid curve is  
the linear fit to our Monte Carlo estimates. The dashed curve represents 
the scaling behaviour predicted by Mack and G\"{o}pfert (eq.\ref{eqn06}) 
and the 
dash-dot curve is the strong coupling result of eq. (\ref{eqn08}).} 
\label{fig8}
\end{figure}

Figs. \ref{fig8} shows the behaviour of the fitted value of the string 
tension $K (=a^{2}\sigma )$
as a function of $\beta$ for the isotropic case (
$\Delta \tau =1$), plotted in comparison to 
results predicted for string tension using the large-$\beta$ approximation 
(eq. \ref{eqn06}). We also include the strong-coupling values and show the  
leading order strong-coupling prediction (eq. \ref{eqn07}). As can be seen 
that our 
estimates of  the string 
tension, in the weak-coupling region  are 
much larger than the Villain prediction, though slopes are more or less 
well 
determined. However one does not expect agreement of precise magnitude, 
since our estimates are obtained using Wilson action, and not the Villian 
approximation. But one does expect that the slope of the string tension in 
both cases should ultimately be the same as one approaches the continuum 
limit.  Mack and G\"{o}pfert \cite{gop82} argued that eq. (\ref{eqn06}) 
represents 
the true 
asymptotic behaviour of the string tension, apart from the value of the 
renormalization
constant  $c$. Such a behaviour was obtained by a semi-classical treatment 
of 
the theory with the effective action without the correction terms. Thus it 
is possible that eq. (\ref{eqn06}), including the normalization constant, 
becomes exact as $1/\beta \rightarrow 0$, if one estimates the actual 
value of $c$.  
We estimated the actual value of $c$ by graphing the $\mbox{e}^{
2.494\beta}\times 4.443K$
against $1/\beta$ and extrapolating to $1/\beta \rightarrow 0$. Using 
estimated value of the renormalization constant, $c =44$, we plotted the 
asymptotic form of the string tension in weak-coupling limit in comparison 
to our estimates. The result agrees very well with our 
estimates.\\

\begin{figure}
\centerline{\psfig{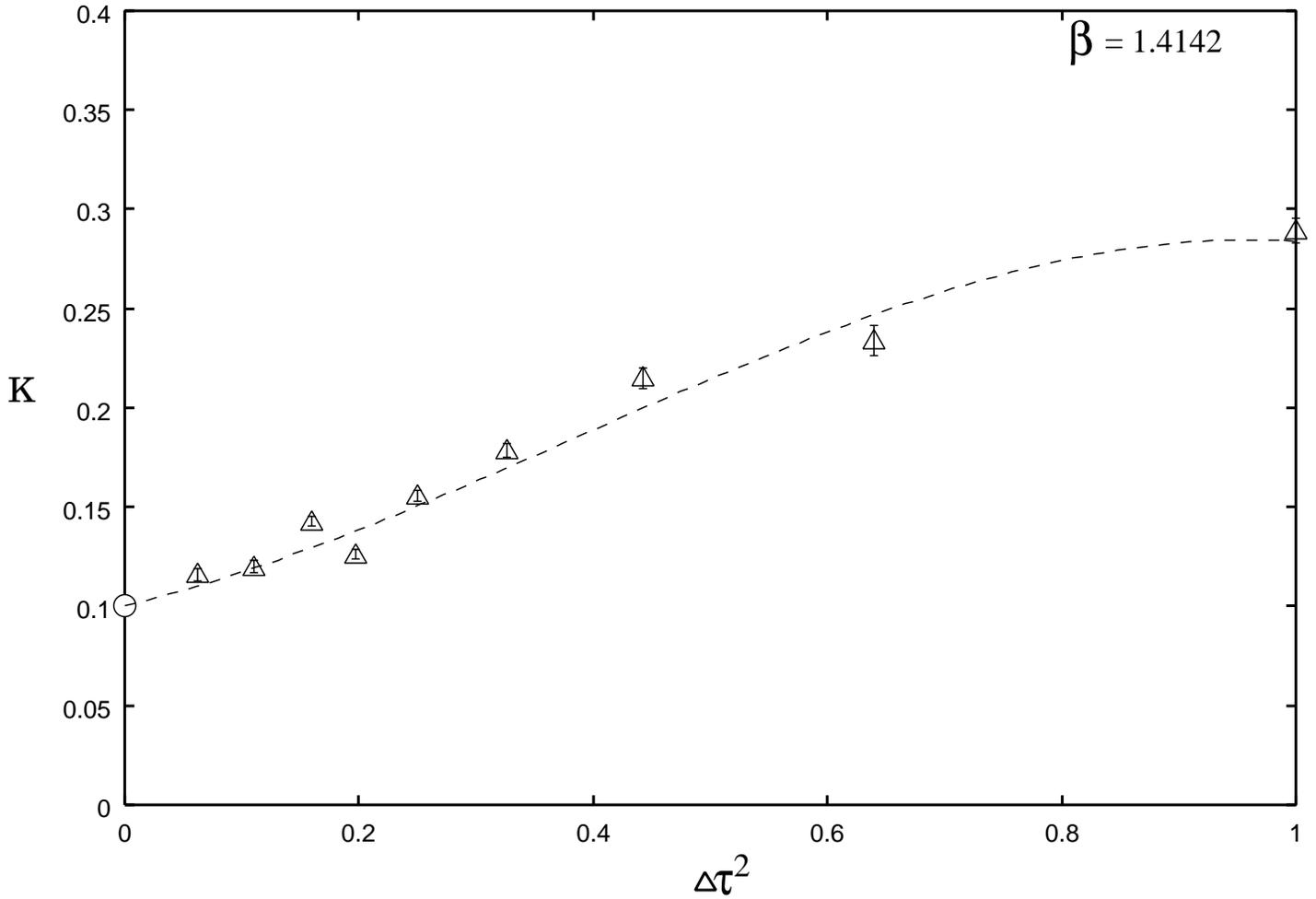}}
\caption{String tension against $\Delta \tau^{2}$ at $\beta =1.4142$. 
Extrapolation to Hamiltonian limit is performed by a cubic fit shown by 
the dashed line and the 
earlier 
Hamiltonian estimate [27] is  shown by $\bigcirc$.}
\label{fig9}
\end{figure}

\begin{figure}
\centerline{\psfig{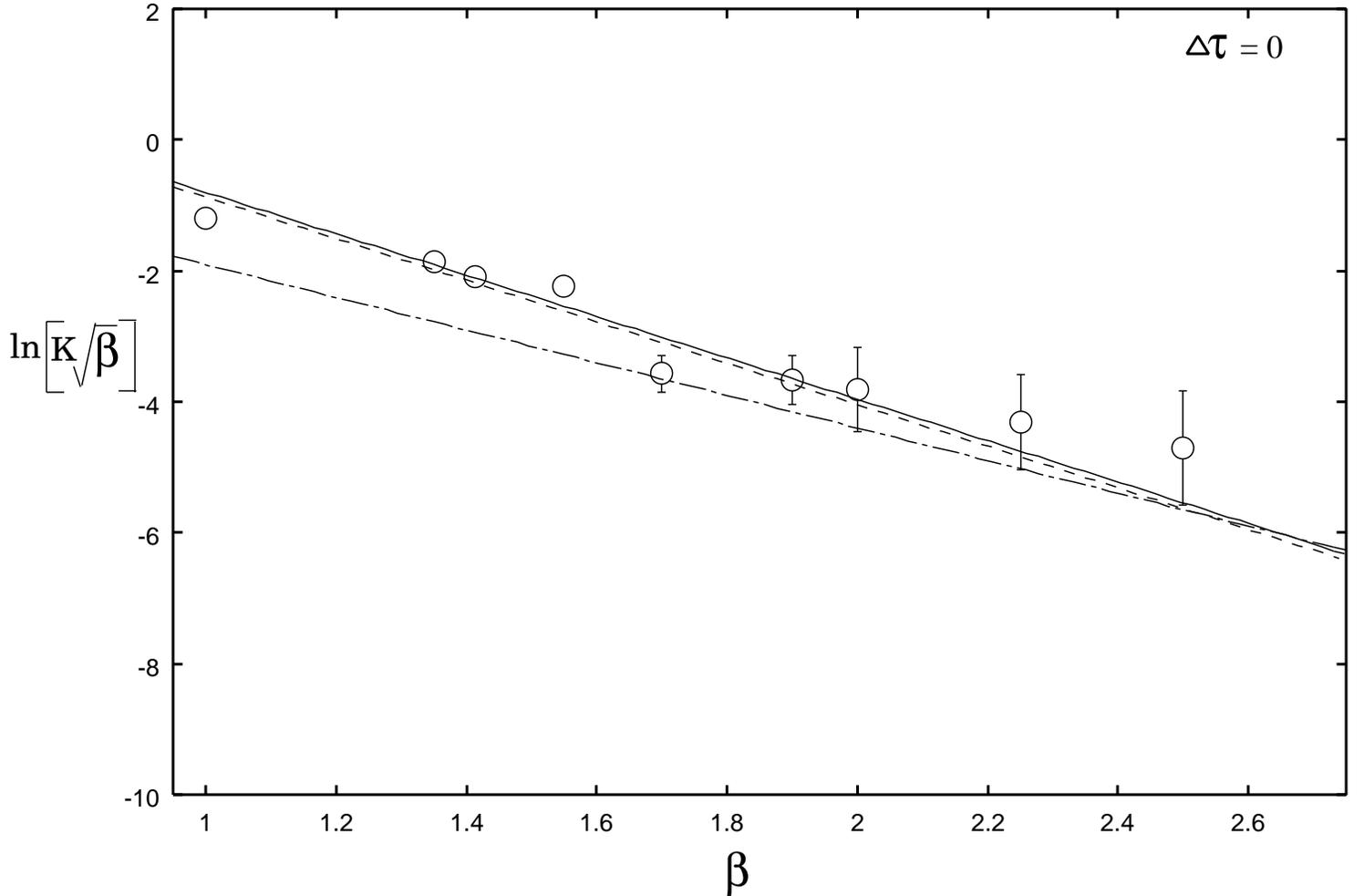}}
\caption{plot showing the estimates of string tension in the Hamiltonian limit
as a function of  $\beta$. 
The solid line is the weak-coupling prediction with estimated value of the 
renormalization constant $c$, 
and dashed line 
is the linear fit to our Monte Carlo data. The dash-dot curve represents
 the large-$\beta$ prediction of  Mack and Gopfert 
 using Villian approximation. }
\label{fig10}
\end{figure}

Fig. \ref{fig9} shows the behaviour of the string tension ($K$)
as a function of the anisotropy ${\Delta \tau}^{2}$, for fixed
coupling $\beta = \sqrt{2}$.
An extrapolation to the Hamiltonian limit $\Dtau \rightarrow 0$ is
performed by a simple cubic fit. Again the extrapolation agrees rather
well with earlier Hamiltonian estimates\cite{ham94}. Note that this
quantity depends rather strongly on $\Dtau$: there is a factor of three
difference between the values at $\Dtau = 0$ and $\Dtau = 1$.
Extrapolating to $\Dtau \rightarrow 0$, estimates of the string tension in 
the Hamiltonian limit, for various $\beta$ values 
are obtained. The plot of these estimates, in comparison to the 
large-$\beta$ prediction of eq. (\ref{eqn06}) is shown 
in Fig. \ref{fig10}. As can be seen, the weak-coupling results, with 
estimated value for constant coefficient, agree 
very well with our estimates. The comparison with 
the theoretical prediction using Villian form is 
not bad either, especially in the large-$\beta$ region. In this region,
upto $\beta \approx 1.8$, the results agree well within errors  
and slopes are very well defined.

\begin{figure}
\centerline{\psfig{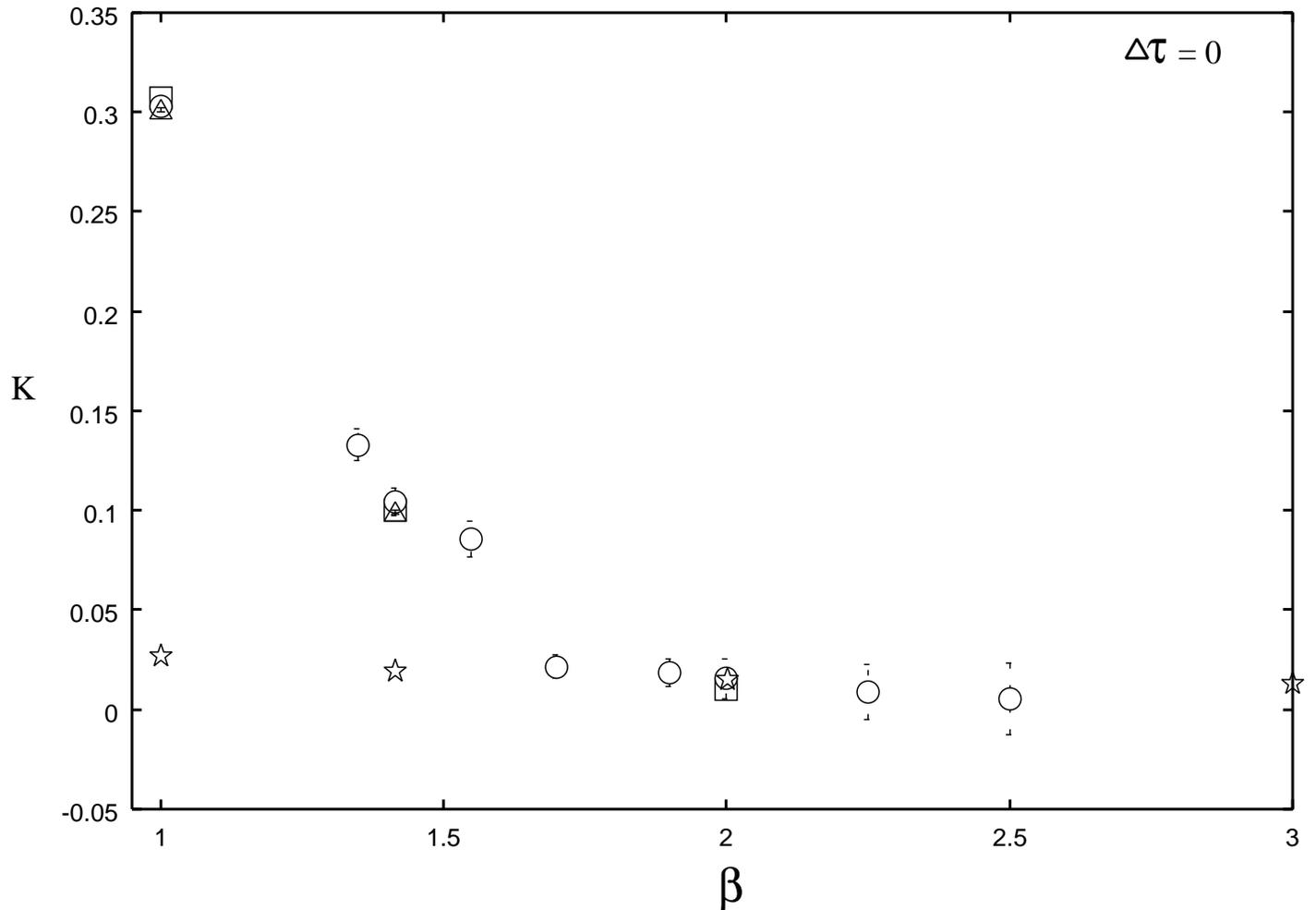}}
\caption{Hamiltonian estimates of string tension as a function of 
$\beta$. 
Our estimates of string tension are labeled by
$\bigcirc$ symbol. Earlier results from series 
expansion [8], Greens function Monte Carlo [17], and finite size scaling 
[37] 
are labeled by $\triangle$, $\Box$ and $\star$ respectively.  
As is seen that our estimates make a very good contact with earlier 
Hamiltonian estimates obtained by many other techniques}
\label{fig11} 
\end{figure}

Fig. \ref{fig11} shows the plot of Hamiltonian estimates as 
a function of $\beta$, plotted in comparison to the previous estimates 
obtained from strong-coupling [8], Greens function Monte Carlo 
\cite{ham00} 
and finite-size scaling \cite{ham93} 
The agreement with the series expansion and GFMC
results is once again excellent. The  estimates of finite-size scaling 
describe the data at $\beta=2.0$.

\section{Summary and conclusion}
In this paper, we used numerical simulation 
techniques on anisotropic lattices for U(1) model in (2+1) dimensions to 
obtain a clear picture of the confinement and improve our knowledge of the 
string tension. 
Our first aim was to implement Path Integral  Monte Carlo method to the
U(1) model, which is one of the simplest models and has been studied by 
various other methods. It seems to work extremely well 
for the mean 
plaquette in the strong-coupling and weak-coupling regimes. The simulation 
results were extrapolated to $\Dtau \rightarrow 0$, and Hamiltonian 
estimates for mean plaquette were obtained. A comparison was made with the 
previous estimates obtained in this limit and agreement was excellant.    
The second  aim was the get a better picture of the confinement and the  
numerical study of the 
asymptotic behaviour of the string tension in the large-$\beta$ limit.
It was shown that at large separations the U(1) theory in (2+1) dimensions 
 shows a linear behaviour as predicted by Polyakov. However, 
 the presumed  logarithmic term could not be tested as our data do not 
extend to very small separations.
The estimates of string tension have confirmed very well the 
weak-coupling predictions, 
provided that one takes the  estimated  value of the constant 
coefficient $c$ rather than its value in the classical approximation 
using effective action without correction terms. For 
$\beta <2.0$, the data is in good agreement with
$O(\beta^{2})$ strong-coupling  predictions. Once again our Hamiltonian 
estimates for the string tension fit very well with the series and GFMC 
results.

This seems to suggest that 
PIMC does work better than other methods, as stated by Ceperley. 
The first clear picture of the static quark potential has been
obtained, showing clear evidence of the linear confining potential at
large distance predicted by Polyakov. Our results from the PIMC do appear 
to be in agreement with results of the series expansions.

\section*{Acknowledgments} This work was supported by the Australian
Research Council. We are grateful for access to the
computing facilities of the Australian Centre for Advanced Computing and
Communications (ac3) and the Australian Partnership for Advanced
Computing (APAC).

\appendix

\section{Weak-coupling prediction using Villian approximation}

Here we extend the discussion of the weak-coupling behaviour of the mass
gap to the anisotropic case, following the discussions of Banks, Myerson
and Kogut\cite{ban77} and Ben-Menahem\cite{ben79}.
This can be probed by means of the expectation value
of the Wilson loop,  $<W(C)>$, which for pure gauge theory is given by
\begin{equation}
<W(C)> = \frac{Z(J)}{Z(0)}
\label{eqnA1}
\end{equation}
where
\begin{equation}
Z(J) = \prod_{r,\mu}\bigg\{ \int d\theta_{\mu}(r)\bigg\}\mbox{exp}\left[
-S-i\sum_{r,\mu}J_{\mu}(r)\theta_{\mu}(r)\right]
\label{eqnA2}
\end{equation}
and $J_{\mu}(r)$ is a vector field associated with a rectangular contour C
of width R and length T.
The extra piece of action
associated with the external charge is
\begin{displaymath}
S_{J} = \int d^{3}rJ_{\mu}(r)\theta_{\mu}(r)
\label{eqnA3}
\end{displaymath}
For a point charge, $S_{J}$ is given by the path integral of the gauge
field along the world line of the charge.
Therefore we choose
\begin{equation}
J_{\mu}(r)   = \left\{ \begin{array}{ll}
  1 & \textrm{if $r\rightarrow r+\hat{\mu}$ is on C} \\
 -1 &  \textrm{if $r+\hat{\mu}\rightarrow $ is on C} \\
  0 & \textrm{otherwise}
   \end{array}
 \right.
\label{eqnA4}
\end{equation}
and this current is conserved: $\Delta_{\mu}J_{\mu}(r)=0$. We will
assume a spacelike Wilson loop for this discussion.

It is convenient to work in the temporal gauge
\begin{equation}
\theta_t(\vec{r}) = 0
\label{eqnA5}
\end{equation}
Then one can separate the angular variable $\theta_{i}(r)$ into longitudinal
and transverse parts,
\begin{equation}
\theta_{i}(r) = \triangle_{i}\rho (r) -\epsilon_{ij}
\frac{1}{\triangle_{1}^{2}+\triangle_{2}^{2}}
\triangle_{j}\theta (r)
\label{eqnA6}
\end{equation}
whence it follows that $\theta_{ij}(r) = \theta (r)$.
In the temporal gauge
eq. ({\ref{eqnA2}}) can be written as
\begin{equation}
Z(J) = \prod_{r}\int_{-\infty}^{\infty}\{d\theta (r)\}\mbox{exp}\left[
-\beta \Dtau \sum_{r}\bigg(1-\cos \theta (r)\bigg)-\frac{\beta}{2\Dtau}
\sum_{r}\theta (r)
\frac{\bar{\triangle}_{t}\triangle_{t}}{\bar{\triangle}_{i}\triangle_{i}}
\theta (r) +i\sum_{r}J_{i}\epsilon_{ij}\frac{1}{\triangle_{k}^{2}}
\triangle_{j}\theta (r)\right]
\label{eqnA8}
\end{equation}
with $\triangle_{i}^{2} = \triangle_{1}^{2}+\triangle_{2}^{2}$. 
 $\bar{\triangle}$ and
${\triangle}$  are defined are the lattice difference operators.
To calculate the path integral we replace the potential term by its Villain
form
\begin{equation}
 \mbox{e}^{-\beta (1-\cos\theta(r))} \rightarrow
\sum_{l=-\infty}^{\infty}\mbox{e}^{il\theta}\mbox{e}^{-\beta}I_{l}(\beta).
\label{eqnA9}
  \end{equation}
  where $I_{\l}(\beta )$ is the modified Bessel function and
$\mbox{exp}(il\theta )$ is the character of the plaquette
  variable in the $l$th irreducible representation of the compact U(1)
group.

For large $\beta$ the modified Bessel function can be approximated by
\begin{equation}
 {e}^{-\beta}I_{l}(\beta) \approx \frac{1}{\sqrt{2\pi \beta}}
\mbox{e}^{-l^{2}/2\beta}
\label{eqnA10}
\end{equation}
The Gaussian integration over $\theta (r)$ reduces the path integral to
\begin{eqnarray}
Z(J) & \propto & \prod_{r}\sum_{l(r)=-\infty}^{\infty}
\mbox{exp}\left[-\frac{1}{2\beta \Dtau }\sum_{r}l(r)\bigg(\Dtau^{2}
\frac{\bar{\triangle}_{i}\triangle_{i}}{
\bar{\triangle}_{t}\triangle_{t}}+1\bigg)l(r)-
\frac{\Dtau}{\beta}\sum_{r}l(r)\frac{1}{\bar{\triangle}_{t}\triangle_{t}}
\epsilon_{ij}J_{i}\triangle_{j}
\right.
\nonumber\\
& & \left. - \frac{\Dtau}{2\beta}\sum_{r}J_{i}(r)
\frac{1}{\bar{\triangle}_{t}\triangle_{t}}J_{i}(r)\right]
\label{eqnA11}
\end{eqnarray}
The convergence of the sum over $l$ can be improved by making use of
the Poisson
 summation formula,
\begin{equation}
\sum_{l=-\infty}^{\infty}h(l) = \sum_{m=-\infty}^{\infty}\int d\phi
h(\phi)\mbox{e}^{2i\pi mr}
\label{eqnA12}
\end{equation}
where $h$ is an arbitrary function.

Substituting $l(r) = \triangle_{t}L(r)$ and using eq. (\ref{eqnA10}), 
Eq. (\ref{eqnA11}) becomes
\begin{eqnarray}
 Z(J) & \propto & \prod_{r}\int_{-\infty}^{\infty}d\phi (r)
\sum_{m(r)=-\infty}^{\infty}\left[\mbox{exp}\bigg\{-\frac{1}{2\beta \Dtau}
\sum_{r}\phi (r)\bigg(\Dtau^{2}\bar{\triangle}_{i}\triangle_{i}
+\bar{\triangle}_{t}\triangle_{t}\bigg)\phi (r)
\right.
\nonumber\\
& & \left.
 + \frac{\Dtau}{\beta }\sum_{r}\phi (r)
 \frac{1}{\triangle_{t}}\epsilon_{ij}J_{i}\triangle_{j}
 +2i\pi \sum_{r}m(r)\phi (r)\bigg\} \right] 
\nonumber\\
& & 
\times
 \mbox{exp}\left[-\frac{\Dtau}{2\beta }
\sum_{r}J_{i}(r)\frac{1}{\bar{\triangle}_{t}\triangle_{t}}J_{i}(r)\right]
\label{eqnA13}
\end{eqnarray}
Finally doing the  Gaussian integral over $\phi$ yields;
\begin{eqnarray}
Z(J) & \propto & \prod_{r}\sum_{m(r)=-\infty}^{\infty}\mbox{exp}\left[
-2\pi^{2}\beta \Dtau \sum_{r} m(r)
\bigg(\frac{1}{\Dtau^{2}\bar{\triangle}_{i}\triangle_{i}+
\bar{\triangle}_{t}\triangle_{t}}\bigg)m(r)
\right.
\nonumber\\
& & \left.
+2\pi i\Dtau^2\sum_{r}m(r)\bigg(\frac{1}{\Dtau^{2}\bar{\triangle}_{i}\triangle_{i}+
\bar{\triangle}_{t}\triangle_{t}}\bigg)
\frac{1}{\triangle_{t}}\epsilon_{ij}J_{i}\triangle_{j}\right]
\nonumber\\
& & \times \mbox{exp}\left[-\frac{\Dtau}{2\beta }\sum_{r,i}J_{i}
\frac{1}{\Dtau^{2}\bar{\triangle}_{i}\triangle_{i}+
\bar{\triangle}_{t}\triangle_{t}}J_{i}(r)\right]
\label{eqnA14}
\end{eqnarray}
The partition function, $Z(J=0)$, is given by
\begin{equation}
Z(J=0)  =  \prod_{r}
\sum_{m(r)=-\infty}^{\infty}\mbox{exp}\left[-2\pi^{2}\beta
\sum_{r,r'} m(r)G(r-r')m(r')\right]
\label{eqnA15}
\end{equation}
where
\begin{equation}
G(r-r') = \frac{\Dtau\delta_{rr'}}{\Dtau^{2}\bar{\triangle}_{i}\triangle_{i}+
\bar{\triangle}_{t}\triangle_{t}}
\label{eqnA16}
\end{equation}
 is a  three-dimensional  lattice massless propagator.

The propagator can be expressed as a momentum integral:
\begin{equation}
G(r,t) = \int_{-\pi}^{\pi}\frac{dk_{0}}{2\pi}\int_{-\pi}^{\pi}
\frac{dk_{1}}{2\pi}
\int_{-\pi}^{\pi}\frac{dk_{2}}{2\pi}\frac{\Dtau\mbox{e}^{ik.
r+k_{0}t)}}{4\sin^{2}k_{0}/2+\Dtau^{2}\bigg(
4-2\cos k_{1}-2\cos k_{2}\bigg)}
\label{eqnA17}
\end{equation}
For the isotropic lattice ($\Dtau = 1$), $G(k=0) = 0.2527$; while in the
limit $\Dtau \rightarrow \infty$, $G(k=0)$ takes the value $0.3214$.\\
The resulting partition function may be written as
\begin{equation}
Z(J) = Z_{spin-wave}Z_{ext}Z_{monopoles}
\label{eqnA18}
\end{equation}
where
\begin{eqnarray}
Z_{spin-wave} & = & \int [d\chi
]\mbox{exp}\left[-\frac{1}{\beta}\sum_{r,i}
\bigg(\triangle_{i}\chi (r)\bigg)^{2}\right]\nonumber\\
Z_{ext} & =& \mbox{exp}\left[-\frac{1}{2\beta}\sum_{r,r',i}
J_{i}(r)G(r-r')J_{i}(r')\right]\nonumber\\
Z_{monopole} &= & \sum_{r}\sum_{m(r)=-\infty}^{\infty}
\mbox{exp}\left[-2\pi^{2}\beta \sum_{r,r'}m(r)G(r-r')m(r')
+2i\pi \sum_{r,r'}G(r-r')\eta (r)m(r)\right]
\label{eqnA19}
\end{eqnarray}
The first factor in eq. (\ref{eqnA17}), the ``spin-wave" piece, 
contributes the ordinary
Coulomb part to the potential between the external charges. The rest of
the eq. (\ref{eqnA17}) is the partition sum of the
Coulomb gas of magnetic
 monopoles  interacting with a
stationary current loo via $G(r-r^{'})$. This three-dimensional Coulomb 
gas
is always in a plasma phase. The only thing that changes with $\beta$ is
the density of the monopoles which decays exponentially to zero as $\beta$
increases.  The monopole density generated by the external loop and
coupling to the lattice monopoles m(r) is a dipole sheet along the surface
spanning the contour C. Thus eq. 
(\ref{eqnA17}) represents a dipole sheet immersed in
a monopole gas, which interacts directly with the dipole sheet. The gas
becomes polarized, and the monopoles of opposite polarity accumulate on
the two sides of the sheet, screening both the magnetic field away from
the sheet and the monopole-monopole interaction, 
thus minimizing the energy of the system.
This gives the
correlation function a finite range, whose inverse is interpreted as a
mass of the scalar field.\\
In the naive continuum limit, the Coulomb self-energy of the monopoles   
blows up  and their density goes to zero exponentially. This causes the
coefficient of the linear force law to go to zero and one regains the
usual continuum theory, i.e., free electromagnetism in three dimensions; 
but if
one renormalizes or rescales in the standard way so as to maintain the 
mass gap constant, then one obtains a confining theory of free massive 
bosons.
The detailed analysis of  Polyakov\cite{pol78} shows that the  
sheet cannot
be screened completely and there is a finite energy left that binds the   
quarks through a potential which grows linearly with separation.
Polyakov  showed that for arbitrarily large finite $\beta$,
theory has
 a mass gap and there appears a linear potential appears between static
charges due to
instantons in the lattice theory.\\
Splitting the exponent in eq. (\ref{eqnA17}) into pieces with $r=r'$ and
$r\neq r'$, the monopole partition sum can be written as
\begin{eqnarray}
Z_{monopole}(J) & = & \prod_{r}\int d\phi 
(r)\mbox{exp}\left[-\frac{1}{8\pi^{2}\beta}\sum_{r}\bigg(\Delta_{i}\phi 
(r)\bigg)^{2}
-\sum_{m=-\infty}^{\infty}2\pi^{2}\beta G_{0}
\right.
\nonumber\\
& & \left. - 
2i\pi\sum_{r}\eta (r)m(r)+i\sum_{r}\phi (r)m(r)\right]
\label{eqnA20}
\end{eqnarray}
 For large $\beta$,
one can neglect all the monopoles with charge other than
0, $\pm 1$ in the sum,
since the monopoles with a higher charge gets extra
powers of $\mbox{e}^{-\beta}$. \footnote{
This argument becomes
problematical when $\Dtau \rightarrow 0$ \cite{ben79}. In this limit the 
monopoles of higher charges contribute to the action and hence are no 
longer suppressed.}
Applying the large-$\beta$ approximation to the monopole partition
function, the above equation can be written, in the naive continuum limit, 
as
\begin{equation}
Z_{monopole}(J)  \approx  \int D\phi 
\mbox{exp}\left[-\frac{g^{2}}{4\pi^{2}}\int 
\bigg\{\frac{1}{2}(\partial_{i}\phi )^{2}-M^{2}\cos\bigg(\phi +2\pi 
\eta\bigg)\bigg\}d^{3}x\right] 
\label{eqnA21}
\end{equation}
where the lattice photon mass $M$ is given by
\begin{equation}
M^{2} = \frac{8\pi^{2}}{g^{2}a^{3}}\mbox{e}^{-2\pi^{2}G_{0}/g^{2}a}.
\label{eqnA23}
\end{equation}
At this point out analysis parallels the work of original work of Banks 
et al \cite{ban77}.\\
The the expectation value of the Wilson loop becomes
\begin{eqnarray}
<W(C)> & \approx &
\mbox{exp}\left[-\frac{1}{2\beta}\sum_{r,r',i}J_{i}(r)
G(r-r')J_{i}(r')\right]
\prod_{r}\int d\phi
(r)\mbox{exp}\left[-\frac{1}{8\pi^{2}\beta}\sum_{r}\bigg(\Delta_{i}\phi
(r)\bigg)^{2}
\right.
\nonumber\\
& & \left.
-2\mbox{e}^{-2\pi^{2}\beta G_{0}}\sum_{r}\bigg(1-\cos (\phi (r)+
2\pi \eta (r))\bigg)\right]
\label{eqnA24}  
\end{eqnarray}

\section{Large-$\beta$ prediction in Hamiltonian formulation}

We now do the Hamiltonian study of U(1) model in (2+1) dimensions using
weak coupling perturbation theory. The Hamiltonian formulation of compact
U(1) gauge theory has been studied by Banks et al. \cite{ban77} 
and Drell et al. \cite{dre79} following the Kogut and Susskind 
approach \cite{kog75}. \\
Consider the anisotropic Wilson gauge action for U(1) model in (2+1)
dimensions,
\begin{equation}
S = \beta \left[\frac{1}{\Dtau }\sum_{i}P_{it} +\Dtau
\sum_{i<j}P_{ij}\right]
\label{eqnB01}
\end{equation}
where we made assumed the lattice symmetric in spatial directions with
$a_{s}=1$ and $a_{t}=\Dtau$.\\
The plaquette variable is given by
\begin{displaymath}
P_{ij} = 1-\cos\theta_{ij}^{P}
\end{displaymath}    
and $\beta =1/g^{2}$.

In the weak coupling approximation
\begin{equation}
\cos \theta_{ij}^{P} \simeq 1-\frac{1}{2}{\theta_{ij}^{P}}^{2}
\label{eqnB02}
\end{equation}
The plaquette angles $\theta^{P}$ are related to the link angles $A_{l}$
by
\begin{equation}     
\theta^{P}(\tilde{n}^{'},\tilde{\hat{k}}) =
\epsilon_{ijk}\left[A_{l}(\b{n},\hat{i})
-A_{l}(\tilde{n}+\hat{j},\hat{i})+A_{l}(\tilde{n}+\hat{i},\hat{j})
A_{l}(\tilde{n},\hat{j})\right]
\label{eqnB03}
\end{equation}
where $(\tilde{n}^{'},\tilde{\hat{k}})$ denote a plaquette centered at
$\tilde{n}^{'}$
and oriented in the $\hat{k}$ while $(\tilde{n}^{'},\hat{i})$ describes a 
link
that starts at site $\tilde{n}$ in the direction $\hat{i}$, with
\begin{displaymath}    
\tilde{n}^{'} = \tilde{n} +(\frac{\hat{i}+\hat{j}}{2})
\end{displaymath}
The Fourier transformations of the plaquette angles and the link angle
give
\begin{eqnarray}
\theta_{\tilde{k}}(\hat{\mu}) & = &  
\frac{1}{\sqrt{N}}\sum_{\tilde{n}^{'}}
\mbox{e}^{-i\tilde{k}\cdot
\tilde{n}^{'}}\theta_{P}(\tilde{n}^{'},\hat{\mu})\nonumber\\
A_{\tilde{k}}q(\hat{\mu}) & = &  \frac{1}{\sqrt{N}}\sum_{\tilde{n}}
\mbox{e}^{-i\tilde{k}.\tilde{n}}A_{l}(\tilde{n},\hat{\mu})      
\label{eqnB04}
\end{eqnarray}
where $N=N_{s}^{2}N_{t}$ is the total number of sites; then
\begin{eqnarray}
A_{\tilde{k}}^{\dagger}(\hat{\mu}) & = & A_{-\tilde{k}}(\hat{\mu}) 
\nonumber\\
\theta_{\tilde{k}}^{\dagger}(\hat{\mu}) & = &  
\theta_{-\tilde{k}}(\hat{\mu})
\label{eqnB05}
\end{eqnarray}
Therefore eq. (\ref{eqnB03}) can be written as
\begin{equation}
\theta_{\tilde{k}}(\hat{\mu}) = \sum_{\hat{\nu}}G_{\tilde{k}}
(\hat{\mu},\hat{\nu})     
A_{\tilde{k}}(\hat{\nu})
\label{eqnB06}
\end{equation}
where
\begin{equation}
G_{\b{k}}(\hat{\mu},\hat{\nu}) = 2i\sum_{\nu, \rho}\epsilon_{\mu\nu\sigma}
\mbox{e}^{-ik_{\nu}/2}\sin (k_{\sigma}/2)
\label{eqnB07}
\end{equation}
Choosing the temporal gauge $A_{0} = 0$, the time-like plaquette
contributions, in the weak coupling approximation, become:
\begin{eqnarray}   
\sum_{\tilde{n}}\theta_{i0}^{2}(\tilde{n}) & = &
\sum_{\tilde{k}}\theta_{i0,\tilde{k}}^{\dagger}
\theta_{i0,\tilde{k}}\nonumber\\
 & = & \sum_{\tilde{k}}A_{\tilde{k}}^{\dagger}(\hat{i})
A_{\tilde{k}}(\hat{i})\times 4\sin^{2}(k_{0}/2)
\label{eqnB08}
\end{eqnarray}
while the space-like plaquette contribution becomes:        
\begin{eqnarray}
\sum_{\tilde{n}}\theta_{12}^{2}(\tilde{n}) & = &
\sum_{\tilde{k}}\theta_{12,\tilde{k}}^{\dagger}\theta_{12,\tilde{k}}
\nonumber\\
 & = & \sum_{\tilde{k}}A_{\tilde{k}}^{\dagger}R_{\tilde{k}}A_{\tilde{k}}
\label{eqnB09}
\end{eqnarray}
where $A_{\tilde{k}}$ and $R_{\tilde{k}}$ are the vectors and matrices,
 $(2\times 2)$ in $(\hat{1},\hat{2})$ square, respectively, and
$R_{\tilde{k}} =G_{\tilde{k}}^{\dagger}G_{\tilde{k}}$. Therefore
\begin{equation}
R_{\tilde{k}} = \left( \begin{array}{ccc}
\sin^{2}(k_{y}/2) & -\mbox{e}^{i(k_{x}/2 -k_{y}/2)}\sin k_{x}/2 \sin 
k_{y}/2 \\
-\mbox{e}^{i(k_{y}/2 -k_{x}/2)}\sin k_{x}/2 \sin k_{y}/2 &   
\sin^{2}(k_{x}/2)
\end{array} \right)
\label{eqnB10}
\end{equation}
The matrix $R_{\tilde{k}}$ is hermitian, and has eigenvalues:
\begin{eqnarray}
\lambda_{\tilde{k}}(1) & = & 4\sum_{i=1}^{2}\sin^{2}(k_{i}/2) \nonumber\\
\lambda_{\tilde{k}}(1) & = & 0
\label{eqnB11}
\end{eqnarray}
We can define a unitary transformation  
\begin{equation}
\omega_{\tilde{k}} = u_{\tilde{k}}A_{\tilde{k}}
\label{eqnB12}
\end{equation}
which diagonalizes the matrix $R_{\tilde{k}}$:
\begin{equation}
u_{\tilde{k}} = \frac{i}{sqrt{\sin^{2}k_{x}/2+\sin^{2}k_{y}/2}}\left(
\begin{array}{ccc}
-\mbox{e}^{ik_{x}/2}\sin (k_{y}/2) & \mbox{e}^{ik_{y}/2)}\sin k_{x}/2 \\
-\mbox{e}^{ik_{x}/2)}\sin k_{x}/2  &
\mbox{e}^{ik_{y}/2}\sin^{2}(k_{x}/2)  
\end{array} \right)
\label{eqnB13}
\end{equation}
The components of $\omega_{\tilde{k}}$ are
\begin{eqnarray}
\omega_{\tilde{k}}(1) & = &  \frac{i}{c_{1}}
\left[\mbox{e}^{-ik_{y}/2)}\sin k_{x}/2 A_{\b{k}}(\hat{2}) -
\mbox{e}^{ik_{x}/2}\sin (k_{y}/2)A_{\b{k}}(\hat{2})\right]\nonumber\\
\omega_{\tilde{k}}(2) & = &  \frac{i}{c_{1}}
\left[\mbox{e}^{-ik_{x}/2)}\sin k_{x}/2 A_{\b{k}}(\hat{1}) +
\mbox{e}^{ik_{y}/2}\sin (k_{y}/2)A_{\b{k}}(\hat{2})\right] 
\label{eqnB14}
\end{eqnarray}
where $c_{1}=\sqrt{\sin^{2}k_{x}/2+\sin^{2}k_{y}/2}$.\\
The component of $w_{\tilde{k}}$ corresponding to the zero eigenvalue is
\begin{equation}
\tilde{\omega}_{\tilde{k}} = \frac{i}{c_{1}}
\left[(1-\mbox{e}^{-ik_{x}})A_{\tilde{k}}(\hat{x})+
(1-\mbox{e}^{-ik_{y}})A_{\tilde{k}}(\hat{y})\right].
\label{eqnB15}
\end{equation}  
This vanishes at all orders and is an "ignorable" coordinate.\\
Thus the action eq. (\ref{eqnB01}) can be written as
\begin{eqnarray}
S & = & \frac{\beta}{2}\sum_{\tilde{k}}\left[\Dtau \bigg(
\triangle_{\tilde{k}}\omega_{\tilde{k}}^{\dagger}(1)
\omega_{\tilde{k}}(1)\right]\nonumber\\
& & +
\frac{1}{\Dtau}\left[4\sin^{2}(k_{0}/2)\sum_{\mu=1}^{2}
\omega_{\tilde{k}}^{\dagger}(\hat{\mu})\omega_{\tilde{k}}(\hat{\mu})\right]
+ \mbox{const}.
\label{eqnB16}
\end{eqnarray}
where            
\begin{equation}
\triangle_{\tilde{k}} = 4\left[\sin^{2}k_{1}/2 +\sin^{2}k_{2}/2\right]
\label{eqnB17}
\end{equation}
Therefore the partition function can be written as
\begin{equation}
Z = \prod_{\tilde{k}}\{\int d\omega {\tilde{k}}\}\mbox{e}^{-S[\omega ]}
\label{eqnB18}
\end{equation}
where the action can be written in the following form:
\begin{eqnarray}
s[\omega ] & \approx & \frac{\beta}{2}\sum_{\tilde{k}}\left[ 
\omega_{\tilde{k}}^{\dagger}(1)\omega_{\tilde{k}}(1)
\bigg(\Dtau \triangle_{\tilde{k}}+\frac{4}{\Dtau}
\sin^{2}(k_{0}/2)\bigg)
\right.
\nonumber\\
& & \left. +
\omega_{\tilde{k}}^{\dagger}(2)\omega_{\tilde{k}}(2)\bigg(
\frac{4}{\Dtau}
\sin^{2}(k_{0}/2)\bigg)\right]
\label{eqnB19}
\end{eqnarray}
Now $\omega_{\tilde{k}}$ is the ignorable coordinate and will be ignored 
from
here on.\\
The  partition function becomes
\begin{eqnarray}
Z & \approx & \prod_{\b{k}}\left[\frac{2\pi }{\beta}
\frac{1}{\bigg(\Dtau \triangle_{\b{k}}+\frac{4}{\Dtau}
\sin^{2}(k_{0}/2)\bigg)}\right]^{1/2} \nonumber\\
 &  = & \mbox{exp}\left[\frac{N}{2}\mbox{ln}(\frac{2\pi}{\beta}) -
\frac{1}{2}\sum_{\b{k}}\mbox{ln}\bigg(
(\Dtau \triangle_{\b{k}}+\frac{4}{\Dtau}  
\sin^{2}(k_{0}/2)\bigg)\right]
\label{eqnB20}
\end{eqnarray}
This lead to the following free energy density:
\begin{equation}
f =
\frac{1}{2N}\sum_{\tilde{k}}\mbox{ln}\left[\frac{\beta}{2\pi}
d(\tilde{k})\right]
\label{eqnB21}
\end{equation}
where
\begin{equation} 
d(\tilde{k}) = (\Dtau \triangle_{\tilde{k}}+\frac{4}{\Dtau}
\sin^{2}(k_{0}/2)
\label{eqnB22}
\end{equation}
\subsection{Mean space-like plaquette}
At weak coupling the cosine term in the space-like part may be expanded in
the powers of $\theta_{P}$. Therefore
\begin{eqnarray}
<\cos \theta_{ij}> & = & 1 -\frac{1}{2}<\theta_{ij}^{2}> \nonumber\\
 & = & 1-\frac{1}{2}<\frac{1}{N}\sum_{\tilde{k}}\triangle_{\tilde{k}}
\omega_{\tilde{k}}^{\dagger}(1)\omega_{\tilde{k}}(1)>
\label{eqnB23}
\end{eqnarray}
where
\begin{eqnarray}
<\omega_{\tilde{k}}^{\dagger}(1)\omega_{\tilde{k}}(1)>
& = & \frac{\int d\omega_{\tilde{k}}\mbox{e}^{-\frac{\beta}{2}d(\tilde{k})
\omega_{\tilde{k}}^{\dagger}\omega_{\tilde{k}}}
\omega_{\tilde{k}}^{\dagger}\omega_{\tilde{k}}}
{\int d\omega_{\tilde{k}}\mbox{e}^{-\frac{\beta}{2}d(\tilde{k})
\omega_{\tilde{k}}^{\dagger}\omega_{\tilde{k}}}}
\label{eqnB24}
\end{eqnarray}  
Using the integrals
\begin{eqnarray}
\int_{-\infty}^{\infty} d\omega \mbox{e}^{-a\omega^{2}} &  = &
\sqrt{\frac{\pi}{2}} \nonumber\\
\int_{-\infty}^{\infty} d\omega \omega^{2}\mbox{e}^{-a\omega^{2}} & = &
\frac{1}{2a} \sqrt{\frac{\pi}{2}}
\label{eqnB25}
\end{eqnarray}
we get
\begin{equation}  
<\omega_{\tilde{k}}^{\dagger}\omega_{\tilde{k}}> = \frac{1}{\beta 
d(\tilde{k}}
\label{eqnB26}
\end{equation}
Therefore
\begin{equation}
<\cos \theta_{ij}> = 1-\frac{1}{2\beta N}\sum_{\tilde{k}}
\frac{\triangle_{\tilde{k}}}{d(\tilde{k})}
\label{eqnB27}
\end{equation}
Now consider the ``cylindrical limit", $N_{t}\rightarrow \infty$,
$\triangle \tau \rightarrow 0$:
In this limit the sum term in the right hand side of the above
equation becomes;  
\begin{eqnarray}
\sum_{\b{k}}
\frac{\triangle_{\tilde{k}}}{d(\tilde{k})} & \rightarrow &
\sum_{\vec{k}}\triangle (\vec{k})\frac{N_{t}}{2\pi}
\int_{0}^{2\pi}dk_{0}\frac{1}{\triangle \tau +\frac{4}{\triangle \tau}
\sin^{2}(k_{0}/2)}\nonumber\\
& = & \sum_{\vec{k}} \frac{N_{t}}{2\pi}2\int_{0}^{2\pi}
\frac{dx}{\bigg(\triangle \tau \triangle (\vec{k}) +
\frac{x^{2}}{\triangle \tau}\bigg)} \nonumber\\
& = & \sum_{\vec{k}}\frac{N_{t}}{\pi}\frac{\pi}{2\sqrt{\triangle    
(\vec{k})}}
\label{eqnB28}
\end{eqnarray}
In the bulk limit, $N\rightarrow \infty$, this becomes;
\begin{eqnarray}
& \approx & \frac{N_{t}{2}.(\frac{N_{s}}{2\pi})^{2}}\int \int_{0}^{2\pi}
dk_{x}dk_{y}\sqrt{4\bigg(\sin^{2}k_{x}/2+\sin^{2}k_{y}/2}\nonumber\\
& = & \frac{N_{t}}{2}.2N_{s}^{2}C_{1}^{(2)}
\label{eqnB29}
\end{eqnarray}
where
\begin{eqnarray} 
C_{1}^{(2)} &  = &
\frac{1}{N_{s}^{2}}\sum_{\vec{k}}(1-\gamma_{\vec{k}})^{1/2}\nonumber\\
& = & \frac{1}{2N_{s}^{2}}\sum_{\vec{k}}\Delta^{1/2}(\vec{k}) \nonumber\\
& = & 0.958
\label{eqnB30}
\end{eqnarray}
is a standard lattice sum.\\
Therefore
\begin{eqnarray}
<\cos \theta_{ij}> &  = &  1-\frac{1}{2\beta}C_{1}^{(2)} \nonumber\\
& = & 1-\frac{g^{2}}{2}C_{1}^{(2)}\nonumber\\ 
& = & 1-0.479g^{2}
\label{eqnB31}
\end{eqnarray}
This result agrees with the estimates in the Hamiltonian limit \cite{ham96}
Now taking the bulk limit in isotropic case, we get
\begin{equation}
\sum_{\tilde{k}} \frac{\Delta (\vec{k})}{d(\tilde{k})}
 \rightarrow \frac{N}{(2\pi )^{3}}\int \int_{0}^{2\pi}\int
dk_{x}dk_{y}dk_{0}\frac{(\sin^{2}k_{x}/2+\sin^{2}k_{y}/2)}{(
\sin^{2}k_{x}/2+\sin^{2}k_{y}/2+\sin^{2}k_{0}/2)}
\label{eqnB32}
\end{equation}  
By symmetry, we can easily find that
\begin{equation}
\sum_{\tilde{k}} \frac{\Delta (\vec{k})}{d(\tilde{k})} = \frac{2N}{3}
\label{eqnB33}
\end{equation}
Therefore
\begin{equation}
<\cos \theta_{ij}> = 1-\frac{g^{2}}{3}
\label{eqnB34}
\end{equation}
\subsection{Mean time-like plaquette}
In weak coupling approximation the time-like plaquette can be written as    
\begin{eqnarray}
\frac{1}{2}\sum_{i=1}^{2}<\cos \theta_{i0}> & = & 1-\frac{1}{4}
\sum_{i=1}^{2}<\theta_{i0}^{2}>\nonumber\\
& = & 1-\frac{1}{4}<\frac{1}{N}\sum_{\tilde{k}}4\sin^{2}k_{0}/2
A_{\tilde{k}}^{\dagger}(\hat{i})A_{\tilde{k}}(\hat{i})> \nonumber\\
& = & 1-\frac{1}{4N}<\sum_{\tilde{k}}4\sin^{2}k_{0}/2 \bigg(
\omega_{\tilde{k}}^{\dagger}(\hat{1})\omega_{\tilde{k}}(\hat{1})+
\omega_{\tilde{k}}^{\dagger}(\hat{2})\omega_{\tilde{k}}(\hat{2}\bigg)>
\label{eqnB35}
\end{eqnarray}
Now           
\begin{equation}
<\omega_{\tilde{k}}^{\dagger}(\hat{1})\omega_{\tilde{k}}(\hat{1})+
\omega_{\tilde{k}}^{\dagger}(\hat{2})\omega_{\tilde{k}}(\hat{2}\bigg)>
= \frac{1}{\beta d(\tilde{k})}+\frac{\Delta \tau}{4\beta \sin^{2}k_{0}/2}
\label{eqnB36}
\end{equation}
\begin{equation}
\frac{1}{2}\sum_{i=1}^{2}<\cos \theta_{i0}>  = 1-\frac{g^{2}}{4N}
\sum_{\tilde{k}}4\sin^{2}k_{0}/2\left[\frac{1}{d(\tilde{k})}+
\frac{\Delta \tau}{4\sin^{2}k_{0}/2}\right]
\label{eqnB37}
\end{equation}  
The term in the square brackets can be written as
\begin{eqnarray}
\sum_{\tilde{k}}4\sin^{2}k_{0}/2\left[\frac{1}{d(\tilde{k})}+
\frac{\Delta \tau}{4\sin^{2}k_{0}/2}\right] & = & \frac{N}{(2\pi )^{3}}
\int \int_{0}^{2\pi}\int dk_{x}dk_{y}dk_{0} \Delta \sin^{2}k_{0}
\left[\frac{1}{\sin^{2}k_{0}/2}
\right.
\nonumber\\
& & \left.
+\frac{1}{\sin^{2}k_{0}/2 +   
\sin^{2}k_{0}/2+{\Delta \tau}^{2}(\sin^{2}k_{x}/2+\sin^{2}k_{y}/2)}\right]
\nonumber\\
& = & \left\{ \begin{array}{ll}
     0  & (\Delta \tau =0) \\
     N(1+1/3) & (\Delta \tau =1)
      \end{array}
      \right.
\label{eqnB38}
\end{eqnarray}
Thus the mean time-like plaquette gives;
\begin{equation} 
\frac{1}{2}\sum_{i=1}^{2}<\cos \theta_{i0}>  =  \left\{ \begin{array}{ll}
     0  & (\Delta \tau =0) \\
     1-\frac{g^{2}}{3} & (\Delta \tau =1)
      \end{array}
      \right.
\label{eqnB39}
\end{equation}
as is expected.
\subsection{Expectation value of Wilson loop}
The expectation value of the Wilson loop is given by
\begin{equation}  
<W(R,T)> = \frac{1}{Z}\prod_{\tilde{k}}\int d\omega_{\tilde{k}}
W_{c}\mbox{e}^{-S[\omega ]}
\label{eqnB40}
\end{equation}
where the Wilson loop operator is given by
\begin{equation}
W_{C} = \frac{1}{2}\left[\prod_{l\in C}\mbox{e}^{A_{l}}+\mbox{h.c}\right]
\label{eqnB41}
\end{equation}
Taking the separation R in the $x$ direction, for simplicity, the Fourier
transformation of above equation) is
\begin{equation}   
W_{C} =\mbox{exp}\left[ \frac{i}{\sqrt{N}}\sum_{\tilde{k}}
\sum_{\tilde{x},\mu \in  C}
\mbox{e}^{i\tilde{k}.\tilde{x}}A_{\tilde{k}}(\hat{\mu}
\right]+\mbox{h.c}
\label{eqnB42}
\end{equation}
We now consider the contribution from class of ``zero modes" with
$ k =(k_{0},\vec{k}=0)$. These modes are likely to be
responsible for the dominant finite-size corrections 
\cite{ham96}, i.e.,
\begin{equation}  
W_{0}(R,t) =
\mbox{exp}\left[\frac{i}{\sqrt{N}}\sum_{k_{0}}A_{k_{0},\vec{0}}(\hat{x})
R\bigg(\mbox{e}^{ik_{0}x_{01}}-\mbox{e}^{ik_{0}x_{02}}\bigg)\right]
\label{eqnB43}
\end{equation}
Recalling that $A_{0}=0$, the above equation becomes
\begin{eqnarray}
W_{0}(R,t) & = &
\mbox{exp}\left[\frac{2}{\sqrt{N}}\sum_{k_{0}}A_{k_{0},\vec{0}}(\hat{x})
R\mbox{e}^{ik_{0}(\frac{x_{01}+x_{02}}{2})}\sin k_{0}T/2\right]\nonumber\\
& = & 
\mbox{exp}\left[\frac{2}{\sqrt{N}}\sum_{k_{0}}
\omega_{\tilde{k}}(\hat{1})
R\sin k_{0}T/2\right]
\label{eqnB44}
\end{eqnarray}
where $T= x_{02}-x_{01}$ and we have ignored the
$\mbox{e}^{ik_{0}(\frac{x_{01}+x_{02}}{2})}$. Therefore
\begin{equation}
<W_{0}(R,T)> = \frac{1}{Z(0)}\prod_{k_{0}}\int
d\omega_{k_{0}}(\hat{1})\mbox{exp}\left[
\frac{2}{\sqrt{N}}\omega_{k_{0}}R\sin (k_{0}T/2)-\frac{\beta}{2\Delta   
\tau}4\sin^{2}(k_{0}/2)\omega_{k_{0}}^{\dagger}\omega_{k_{0}}\right]
\label{eqnB45}
\end{equation}
Let
\begin{displaymath}
\frac{\beta}{2\Delta
\tau}4\sin^{2}(k_{0}/2)= c(k_{0});
\end{displaymath}
then the numerator of above equation can be written as
\begin{equation}
I_{k_{0}}=\int_{-\infty}^{\infty}d\omega \mbox{exp}\left[
-c(k_{0}\bigg(\omega -i\frac{R\sin (k_{0}T/2)}{\sqrt{N}c(k_{0})}\bigg)^{2} 
\right]\mbox{exp}\left[-
\frac{R^{2}\sin^{2}(k_{0}T/2)}{\sqrt{N}c(k_{0})}\right]
\label{eqnB46}
\end{equation}
Therefore the expectation value of the Wilson loop is given by
\begin{eqnarray}
<W_{0}(R,T)> &  = &  \prod_{k_{0}}\mbox{exp}\left[
\frac{R^{2}\sin^{2}(k_{0}T/2)}{\sqrt{N}2\beta \sin^{2}(k_{0})}\right]
\nonumber\\
&  & \mbox{exp}\left[
-\frac{R^{2}\Delta \tau}{2N_{s}^{2}N_{t}\beta}\frac{N_{t}}{2\pi}
\int_{0}^{2\pi}dk_{0}\frac{\sin^{2}(k_{0}T/2}{\sin^{2}(k_{0}/2}\right]     
\label{eqnB47}
\end{eqnarray}
Now consider the cylindrical limit, $N_{t} \rightarrow \infty$,
$\Delta \tau \rightarrow 0$. In this limit
\begin{equation}
\frac{1}{2\pi}\int_{0}^{2\pi}dk_{0}\frac{\sin^{2}(k_{0}T/2)}{\sin^{2}(k_{0}/2)}
\rightarrow T
\label{eqnB48}
\end{equation}
Therefore
\begin{eqnarray}
<W_{0}(R,T)> & = &  
\mbox{exp}\left[-\frac{g^{2}}{2}(\frac{R}{N_{s}})^{2}
T\Delta \tau \right]\nonumber\\
&  & \mbox{exp}\bigg(-\sigma \times \mbox{area}\bigg)
\label{eqnB49}
\end{eqnarray}
Where
\begin{equation}
\sigma_{eff} = \frac{g^{2}}{2}(\frac{R}{N_{s}^{2}})=
\frac{g^{2}}{2}(\frac{R}{L^{2}})
\label{eqnB50}
\end{equation}
For R=L, i.e., for  pair of Polyakov loops mapping around the lattice in       
the spatial direction, we get
\begin{equation}
\sigma_{eff} = \frac{g^{2}}{2L}
\label{eqnB51}
\end{equation}
This corresponds to the value calculated by Hamer and Zheng 
\cite{ham96}, 
$\sigma = 1/L$, for a string wrapping around the entire lattice, if we
multiply by an extra factor $g^{2}/2$ coming from renormalizing
Hamiltonian. As is obvious that for smaller loops, the effective string
tension is less.

\center
\widetext
\input psfig
\psfull


\begin{references}

\bibitem{cre79}
M. Creutz,
Phys. Rev. Letts. {\bf 43}, 553 (1979)

\bibitem{wil74}
K.G. Wilson,
Phys. Rev. {\bf D10}, 2445 (1974)

\bibitem{kog75}
J. Kogut and L. Susskind,
Phys. Rev. {\bf D11}, 395 (1975)

\bibitem{mor97}
C.J. Morningstar and M. Peardon,
Phys. Rev. {\bf D56}, 4043 (1997)

\bibitem{ham96}
C. J. Hamer, M. Sheppeard, W. H. Zheng and D. Sch\"{u}tte,
Phys. Rev. {\bf D54}, 2395 (1996)

\bibitem{ban77}
T.Banks, S. Raby, L. Susskind, J. Kogut, D.R.T. Jones, P.N. Scharbach
and D.K. Sinclair,
Phys. Rev. {\bf D15}, 1111 (1977)

\bibitem{hor84}
D. Horn and M. Weinstein,
Phys. Rev. {\bf D30}, 1256 (1984)

\bibitem{guo88}
Guo Shuohong, Zheng Weihong and Liu Jiunmin,
Phys. Rev. {\bf D38}, 2591 (1988)

\bibitem{byr02}
T.M.R. Byrnes, P.Sriganesh, R.J. Bursill and C.J. Hamer, to be published
in Phys. Rev. D

\bibitem{bla83}
D.Blankenbecler and R.L. Sugar,
Phys. Rev. {\bf D27}, 1304( 1983)

\bibitem{deg85}
T.A. DeGrand and J. Potvin,
Phys. Rev. {\bf D31}, 871 (1985)

\bibitem{ham94}
C. J. Hamer, K. C. Wang and P. F. Price,
Phys. Rev. {\bf D50}, 4693 (1994)

\bibitem{chi84}
S. A. Chin, J. W. Negele and S. E. Koonin,
Ann. Phys. (N.Y.) {\bf 157}, 140 (1984)

\bibitem{hey83}
D. W. Heys and D. R. Stump,
Phys. Rev. {\bf D28}, 2067 (1983)

\bibitem{kal66}
 Kalos M. H., J. Comp. Phys. {\bf 1}, 257 (1966);
D. M. Ceperley and M. H. Kalos, in {\em Monte Carlo Methods in
Statistical Mechanics}, ed.\ K. Binder (Springer-Verlag, New York,
1979).

\bibitem{sam99}
M. Samaras and C. J. Hamer, Aust. J. Phys. {\bf 52}, 637 (1999)

\bibitem{ham00}
C.J. Hamer, R.J. Bursill and M. Samaras,
Phys. Rev. {\bf D62}, 054511 (2000)

\bibitem{ham00a}
C.J. Hamer, R.J. Bursill and M. Samaras,
Phys. Rev. {\bf D62}, 074506 (2000)

\bibitem{liu74}
K.S.  Liu, M.H. Kalos and G.V. Chester, Phys. Rev. {\bf A10} 303 (1974)

\bibitem{whi79}
P. A. Whitlock, D. M. Ceperley, G. V. Chester and M. H. Kalos,
Phys. Rev. {\bf B19}, 5598 (1979)

\bibitem{amb82}
J. Ambjorn, A.J.G. Hey and S. Otto,
Nucl. Phys. {\bf B210}, 347 (1982)

\bibitem{cod86}
P.D. Coddington, A.J.G. Hey, A.A. Middleton and J.S. Townsend,
Phys. Letts. {\bf B175}, 64 (1986)




\bibitem{chi86}
S. A. Chin, C. Long and D. Robson,
Phys. Rev. Letts. {\bf 57}, 60 (1986),




\bibitem{dre79}
S.D. Drell, H.R. Quinn, vetitsky and M. Weinstein, Phys. Rev. {bf
D19}, 619 (1979)

\bibitem{gro83}
L. Gross,
Commun. Math. Phys. {\bf 92}, 137 (1983)

\bibitem{pol78}
A.M. Polyakov, Phys. Lett. {\bf 72B}, 477 (1978)

\bibitem{gop82}
M. G\"{o}pfert and G. Mack,
Commun. Math. Phys. {\bf 82}, 545 (1982)

\bibitem{ham92}
C. J. Hamer, J. Oitmaa, and Zheng Weihong,
Phys. Rev. {\bf D45}, 4652 (1992)

\bibitem{ham96b}
C.J.Hamer, Zheng Weihong and J. Oitmaa,
Phys. Rev. {\bf D53}, 1429 (1996)

\bibitem{irv83}
A.C. Irving, J.F. Owens and C.J. Hamer,
Phys. Rev. {\bf D28}, 2059 (1983)

\bibitem{hor87}
D. Horn, G. Lana and D. Schreiber,
Phys. Rev. {\bf D36}, 3218 (1987)

\bibitem{mor92}
C.J. Morningstar,
Phys. Rev. {\bf D46}, 824 (1992)

\bibitem{dab91}
A. Dabringhaus, M.L. Ristig and J.W. Clark,
Phys. Rev. {\bf D43}, 1978 (1991)

\bibitem{fan96}
X.Y. Fang, J.M. Liu and S.H. Guo,
Phys. Rev. {\bf D53}, 1523 (1996)

\bibitem{bishop96}
S.J. Baker, R,.F. Bishop and N.J. Davidson, Phys. Rev. {\bf D53}, 2610
(1996).

\bibitem{koo86}
S. E. Koonin, E. A. Umland and M. R. Zirnbauer,
Phys. Rev. {\bf D33}, 1795 (1986)

\bibitem{yun86}
C. M. Yung, C. R. Allton and C. J. Hamer,
Phys. Rev. {\bf D33}, 1795 (1986)

\bibitem{ham93}
C. J. Hamer and Zheng Weihong,
Phys. Rev. {\bf D48}, 4435 (1993)

\bibitem{bat85}
G.G. Batrouni, G.R. Katz, A.S. Kronfeld, G.P. Lepage, B.Svetitsky and
K.G. Wilson, Phys. Rev. {\bf D32}, 2736 (1985)

\bibitem{dav90}
C.T.H. Davies, G.G. Batrouni, G.R. Katz, A.S. Kronfeld, G.P. Lepage, P.
Rossi, B. Svetitsky and K.G. Wilson, Phys. Rev. {\bf D41}, 1953 (1990)

\bibitem{alb87}
M. Albanese {\it et al.}, Phys. Lett. {\bf B192}, 163 (1987)

\bibitem{tep86}
M. Teper, Phys. Lett. {\bf B183}, 345 (1986); K.Ishikawa, A. Sato, G.
Schierholz and M. Teper, Z. Phs. {\bf C21}, 167 (1983).

\bibitem{par83}
G. Parisi, R. Petronzio and F. Rapuano, Phys. Lett. {\bf 128B}, 418
(1983).

\bibitem{tep99}
M.J. Teper, Phys. Rev. {\bf D59}, 014512 (1999)

\bibitem{ban77}
T. Banks, R. Myerson and J. Kogut, Nucl. Phys. {\bf B129}, 493 (1977).

\bibitem{ben79}
S. Ben-Menahem, Phys. Rev. {\bf D20}, 1923 (1979).

\bibitem{kar82}
F. Karsch, Nucl. Phys. {\bf B205}, 285 (1982).

\bibitem{bur88}
G. Burgers, Nucl. Phys. {\bf B304}, 587 (1988).

\bibitem{bra90}
F. Brandstacter {\it{et al}}., Nucl. Phys. {\bf B345}, 709 (1990).

\bibitem{hors81}
R. Horsley and U. Wolff, Phys. Letts. {\bf B105}, 290 (1981).

\bibitem{bhanot80}
G. Bhanot and M. Creutz, Phys. Rev. {\bf D21}, 2892 (1980).

\bibitem{loan}
M. Loan, and C.J. Hamer,
to be published
 
\end{references}
\end{document}